\documentclass[10pt]{article}%
\usepackage{amsfonts}
\usepackage{amsmath}
\usepackage{amssymb}
\usepackage{graphicx}
\usepackage{cite}

\usepackage{color}                    % For creating coloured text and background
\usepackage{hyperref}                 % For creating hyperlinks in cross references

\definecolor{red}{rgb}{1,0,0}           % Standard colours red, green, blue
\definecolor{green}{rgb}{0,1,0}
\definecolor{blue}{rgb}{0,0,1}
\definecolor{Light}{gray}{.80}          % How you can define your own greys
\definecolor{Dark}{gray}{.50}
\definecolor{pink}{rgb}{.95,0.82,0.98}  % How you can define your own colours
\definecolor{yellow}{rgb}{1,1,0}
\definecolor{purple}{rgb}{0.85,0.15,1}

\def \be {\begin{equation}}
\def \ee {\end{equation}}
\def \bea {\begin{eqnarray}}
\def \eea {\end{eqnarray}}
\def \nn {\nonumber}

\def \rr {\raise.35ex\hbox{\small $\prime$}\kern-.17em{\mbox{\large $\imath$}}}
\def \del {\partial}
\def \dels {\partial\kern-.5em / \kern.5em}
\def \As {{A\kern-.5em / \kern.5em}}
\def \Ds {D\kern-.7em / \kern.5em}

\def \a {\alpha}
\def \b {\beta}

\def \d {\delta}
\def \eps {\epsilon}

\def \s {\sigma}

\def \star {\ast}
\newcommand{\vl}{{\vec \ell}}
\newcommand{\vn}{{\vec n}}

\newcommand{\vlam}{{\vec \lambda}}

\begin{document}

%\catcode`\@=11
%\catcode`\@=12
%\twocolumn[\hsize\textwidth\columnwidth\hsize\csname%
%@twocolumnfalse\endcsname

%\begin{narrowtext}

\begin{titlepage}
\thispagestyle{empty}
\begin{flushleft}
%\hfill hep-th/0607052 \\
UT-06-25\hfill January, 2007 \\
\end{flushleft}

\vskip 1.5 cm
\bigskip
\renewcommand{\thefootnote}{\fnsymbol{footnote}}

\begin{center}
\noindent{\LARGE A toy model of open membrane field theory}\\
{\LARGE in constant 3-form flux}
%\vfill
\vskip 2cm
{\large
Pei-Ming Ho\footnote{%
e-mail address: pmho (AT) phys.ntu.edu.tw}}\\
\vskip 3mm
{\it\large
Department of Physics and Center for Theoretical Sciences, \\
National Taiwan University, Taipei 10617, Taiwan, 
R.O.C.}\\
\vskip 5mm
and\\
\vskip 5mm
{\large Yutaka~Matsuo\footnote{e-mail address:
 matsuo (AT) phys.s.u-tokyo.ac.jp}}
\\
\vskip 3mm
{\it\large
Department of Physics, Faculty of Science, University of Tokyo,
Tokyo, Japan\\
\noindent{ \smallskip }\\
}
%\vfill
\vskip 20mm
\bigskip
\end{center}

\begin{abstract}
Based on an explicit computation of the scattering amplitude of 
four open membranes in a constant 3-form background, 
we construct a toy model of the field theory for open membranes
in the large $C$ field limit.
It is a generalization of the noncommutative field theories
which describe open strings in a constant 2-form flux. 
The noncommutativity due to the $B$-field background is now
replaced by a nonassociative triplet product. 
The triplet product satisfies the consistency conditions of lattice 3d gravity, 
which is inherent in the world-volume theory of open membranes. 
We show the UV/IR mixing of the toy model by 
computing some Feynman diagrams. 
Inclusion of the internal degree of freedom
is also possible through the idea of the cubic matrix.
\end{abstract}
\end{titlepage}\vfill\setcounter{footnote}{0} 
\renewcommand{\thefootnote}{\arabic{footnote}} 
\newpage

\setcounter{footnote}{0}
%%%%%%%%%%%%%%%%%%%%%%%%%%%%%%%%%%%%%%%%%%%%%%%%%%%%%%%
%%%%%%%%%%%%%%%%%%%%%%%%%%%%%%%%%%%%%%%%%%%%%%%%%%%%%%%
%%%%%%%%%%%%%%%%%%%%%%%%%%%%%%%%%%%%%%%%%%%%%%%%%%%%%%%
%%%%%%%%%%%%%%%%%%%%%%%%%%%%%%%%%%%%%%%%%%%%%%%%%%%%%%%
%%%%%%%%%%%%%%%%%%%%%%%%%%%%%%%%%%%%%%%%%%%%%%%%%%%%%%%

\section{Introduction}
The formulation and/or the quantization of M(embrane) theory
\cite{deWit:1988ig, Banks:1996vh}
has been one of the long standing issues in string theory.  
There is a variety of difficulties on this problem. 
One of  the conceptually toughest problems may be how to avoid the
instability of the system \cite{deWit:1988ct}. 
%% YM 0305 %%
To cure this problem, one needs
the second quantization of the membrane
from the outset (as emphasized, for example, in \cite{Nicolai:1998ic}).
%% YM 0305 %%
In the infinite momentum frame, BFSS proposal \cite{Banks:1996vh}
is one of the most successful example.  
Nevertheless, since we need to take a particular gauge
and thus covariance in the three dimensional world-volume is lost,
it is clear that we need more efforts.

In the string case, many features of the string dynamics can be
captured by a simpler set-up: quantum
field theory on the noncommutative spacetime
\cite{Seiberg:1999vs, Aoki:1999vr, Minwalla:1999px, Hayakawa:1999yt}.
Since the noncommutativity comes from
the  2-form field $B_{\mu\nu}$
which couples to the string world-sheet, it is natural
to assume the spacetime noncommutativity
represents an essential feature of the string theory.

The corresponding field which couples to the membrane 
world-volume is the 3-form field $C_{\mu\nu\lambda}$.
So we are led to the problem of finding a simpler system
which captures the essential features of the open membrane
dynamics in a constant 3-form field.   
Since the direct quantization of the membrane
is difficult so far, such a simpler set-up is conceptually important.

Since this is a natural development
from the noncommutative spacetime, it has been studied
for a long time by many authors
\cite{Hoppe:1996xp, Awata:1999dz, Xiong:2000gp, Bergshoeff:2000jn, 
Kawamoto:2000zt, Matsuo:2000fh, Sakakibara:2000vx,
Ikeda:2001fq, Kawamura, Pioline:2002ba, 
Curtright:2002fd, Berman:2004jv, Basu:2004ed, Sasakura:2005js, 
Bagger:2006sk}.  So far, unfortunately, 
we do not seem to have a unique or a convergent  
solution which is recognized as the standard choice
by the community. We note that in many literatures where
this issue was studied, the key issues are
(i) ``quantum Nambu bracket'' 
\cite{Nambu:1973qe, Takhtajan:1993vr, Dito:1996xr,
Hoppe:1996xp, Awata:1999dz, Xiong:2000gp, 
Matsuo:2000fh, Sakakibara:2000vx, Kawamura,
Curtright:2002fd}
and/or (ii)``nonassociative geometry''
\cite{Jackiw:1984rd, Cornalba:2001sm, 
Ho:2001qk, Bouwknegt:2004ap, Ellwood:2006my, 
Sasai:2006ua}.  The origin of these concepts
is clear. Nambu bracket
is a natural generalization of Poisson bracket.  While
Poisson structure is defined by 2-form field
which is naturally associated with $B_{\mu\nu}$,
Nambu bracket is defined by higher rank 
anti-symmetric tensors.  As for (ii), the nonassociativity
of the algebra may be identified with 3-cocycle
associated with the constant three form background.
In either case, they are strongly 
correlated to the $C$ field in the membrane theory.
The development in these subjects is very briefly summarized
in the appendix A.

The problems in these subjects are that they are both 
conceptually difficult and at the same 
time there is no clear guideline or principle
which is helpful to convince the community that the solutions
provided are indeed the correct one.
It should be noted, however, that there are already respectable
accumulation of studies where various aspects of these concepts
are proposed.  So the time has already arrived when we can 
consider the application of them to construct
more concrete models of the open membrane.

Toward this goal, we take the following steps to define our model.
\begin{enumerate}
\item We study the free motion of an open membrane
in a constant $C$-field background and observe that the membrane
spreads in directions normal to the momentum
with the area proportional to the momentum and $C$ field.
We also evaluate the four point correlation function
and obtain a phase factor proportional to 
$\sqrt{|\vec p_1\cdot(\vec p_2\times\vec p_3)|}$.
(\S 2--\S3)
\item Based on this observation, we
propose to use the truncated
representation of the open membrane
by a triangle.  This is an analogy of
the noncommutative field theory where
the open string is realized as a straight line between
two points. The product among the membrane fields is defined
by using a tetrahedron whose four faces are identified 
with the triangles associated to the four membrane fields.
Instead of a product defined for two fields,
we have a product for three fields (triplet product).
By combining it with the inner product with another field, 
we obtain the definition of the quartic product.
(\S4)
\item We define a scalar
field theory as a toy model
of the open membrane field theory by using the quartic
product with the phase factor associated with the $C$-field background.
We carry out some explicit computation of the Feynman graphs
and observe the UV/IR mixing which is similar to the
noncommutative field theory \cite{Minwalla:1999px, Hayakawa:1999yt}.
(\S5)
\item The definition of the membrane field based on
the triangle turns out to be  convenient to introduce
 the internal (``color'')  degrees of freedom 
by combining the definition of the cubic matrix theory 
\cite{Ambjorn_book, DePietri:2000ii} 
(see also \cite{Awata:1999dz, Xiong:2000gp, Kawamura}).
We argue that there are $N^3$ degree of freedom 
for each membrane field \cite{N3}.
(\S6)
\item We show that our definition of the membrane field
associated with triangles can be naturally generalized to
those with $n$-gons. As $n$ gets larger, the interactions
among open membranes resemble those of closed
string field theory \cite{Zwiebach:1992ie}.
In this generalized framework, one
can discuss the double dimensional reduction to the
string theory on the noncommutative space 
\cite{Duff:1987bx}. (\S7)
\item After adding some extra rules, the product among the
membrane fields satisfies the consistency conditions of the
3d lattice gravity \cite{r:LatticeGravity}. We argue that
this is the replacement of the associativity in the star product
for the noncommutative case. Appearance of 3d gravity
seems natural in the formulation of the membrane theory.
(\S8)
\item In \S9, we give a brief outllook toward the construction
of the gauge invariant theory.  In the appendix, we give a brief
review  on the nonassociative geometry 
and the Nambu bracket to explain the relation between
previous works and our approach.
\end{enumerate}

\section{Review on noncommutative spacetime}

First we review how a large $B$-field background leads 
to noncommutative spacetime for open strings \cite{NC}. 

For a large $B$-field background in the $X_1$-$X_2$ direction,
the world-sheet action for a fundamental string is dominated by the term
\be
S = \int B dX^1 dX^2 = \int d\tau d\sigma B(\dot{X}_1 X'_2 - \dot{X}_2 X'_1), 
\ee
which is proportional to the world-sheet area.
Since B is large, the path integral is dominated by
the configuration with minimal area.

According to this action, the momentum of an open string is
\be \label{PBX}
P_1 = B\int d\sigma X'_2 = B\Delta X_2,
\qquad
P_2 = -B\int d\sigma X'_1 = -B\Delta X_1.
\ee
Therefore the width of the string is proportional to the momentum,
and the direction of the momentum is perpendicular to the extension of the string
(that is, $P_i \perp X'_i$).

In the low energy limit when oscillation modes can be ignored, 
it is sufficient to approximate the open string by the ground state configuration, 
i.e., a straight line stretched between the endpoints \cite{Zheng,Bigatti}. 

The easiest way to see an indication of noncommutative space is to observe that
the uncertainty relation 
\be \label{UR}
\delta X_1 \delta X_2 \geq \frac{1}{B}
\ee
is satisfied in the following sense. 
For an open string moving in the $X_1$-direction with momentum $P_1$,
we have the usual uncertainty relation 
\be \label{usualUR}
\d X_1 \d P_1 \geq 1
\ee
as well as the new relation
\be
\d P_1 \sim B \d X_2,
\ee
which follows from (\ref{PBX}). 
The uncertainty relation (\ref{UR}) is just a direct result of combining these two relations. 

A more precise way to see the appearance of noncommutative space 
is to check that, 
in the aspect of interactions, 
the noncommutativity of spacial coordinates 
has the same effect as the background $B$ field. 
Hence let us compute the scattering amplitudes of 
$n$ open strings with momenta $k_a$ in the ground state. 
Adding a source term to the action 
\be
S = \int B dX^1 dX^2 + \int d\tau J^i(\tau) X_i(\tau),
\ee
where 
\be
J^i(\tau) = \sum_{a=1}^n k_a^i \delta(\tau - \tau_a), 
\ee
we find the equations of motion
\be
\dot{X}_i = B^{-1}\eps_{ij}\sum_a k_a^j \delta(\tau - \tau_a),
\ee
where $\eps_{ij}$ is the totally antisymmetrized tensor,
for $X_i$ at the world-sheet boundary. 
The solution is 
\be
X_i(\tau) = B^{-1}\eps_{ij}\sum_a k_a^j \theta(\tau-\tau_a) + x^0_i,
\ee
where $\theta(\tau-\tau_a)$ is the step function, 
and $x^0_i$ is a constant of integration. 
Plugging this solution into the action, 
we find the scattering amplitude
%% p %%
% a factor of 1/2 in the exponent here and below %
\be \label{phasefactor} 
e^{iS} = e^{\frac{i}{2}B^{-1}\sum_{a<b} \eps_{ij} k^i_a k^j_b}.
\ee
Because of this phase factor, 
it is convenient to use the notion of noncommutative space 
in the effective field theory.
The Moyal $\ast$ product for the noncommutative space 
\be
[x_i, x_j]_{\ast} = \frac{i}{B}\eps_{ij}
\ee
is defined by 
\be
(f\ast g)(x) = \left. e^{\frac{i}{2}B^{-1}\eps^{ij} \del_i \del'_j} f(x) g(x') \right|_{x' = x} .
\ee
It leads to the same phase factor 
\be
e^{ik_1\cdot x} \ast \cdots \ast e^{ik_n \cdot x} = e^{\frac{i}{2}B^{-1}\sum_{a<b} \eps_{ij} k^i_a k^j_b} e^{i\sum_c k_c \cdot x}
\ee
when multiplying the wave functions of plane waves. 
Therefore the effect of the $B$-field background can be nicely encoded 
in the noncommutativity of space. 

Now consider a Feynman diagram with 3 external legs at tree level.
Each external leg is an open string whose world-sheet is a flat strip 
with straight boundaries. 
The direction of the strip is parallel to the momentum $k$, 
and its width is given by the magnitude of its momentum $|k|$ times $B^{-1}$.
(See Fig. \ref{openstring} (a).)
The scattering amplitude is given by the configuration with minimal area,
which is obviously just to attach the 3 strings on a triangle uniquely 
determined by the momenta $k_a$.
(See Fig. \ref{openstring} (b).)
Note that $B^{-1}$ times the area of the triangle  
gives the exponent of the phase factor (\ref{phasefactor}). 

\begin{figure}[ptb]\label{openstring}
\setlength{\unitlength}{3pt}
\par
\begin{center}
$\begin{array}{cc}
\begin{picture}(40,40)(-20,-20)
% draw triangle
\put(-20,10){\line(1,0){30}}
\put(-20,-10){\line(1,0){30}}
\put(10,-10){\line(0,1){20}}
\put(10,0){\vector(1,0){10}}
\put(15,2){$k$}
\put(0,-4){\vector(0,-1){6}}
\put(0,4){\vector(0,1){6}}
\put(-3,-1){$|k|$}
\end{picture}
&
%\end{center}
%\caption{Fig.\ref{1openstring} An open string in ground state with momentum $k$.}
%\end{figure}
%\begin{figure}[ptb]\label{3openstring}
%\setlength{\unitlength}{3pt}
%\par
%\begin{center}
\begin{picture}(80,80)(-40,-40)
% draw triangle
\put(10,-10){\line(0,1){20}}
\put(10,-10){\line(-3,2){15}}
\put(10,10){\line(-3,-2){15}}
% draw string #1
\put(10,-10){\line(1,0){30}}
\put(10,10){\line(1,0){30}}
\put(40,0){\vector(-1,0){5}}
\put(40,1){$k_1$}
% draw string #2
\put(10,-10){\line(-2,-3){15}}
\put(-5,0){\line(-2,-3){15}}
\put(-13, -27){\vector(2,3){5}}
\put(-15,-24){$k_2$}
% draw string #3
\put(10,10){\line(-2,3){15}}
\put(-5,0){\line(-2,3){15}}
\put(-13, 27){\vector(2,-3){5}}
\put(-15,24){$k_3$}
% caption
%\put(-10,-40){Fig.\ref{3openstring} The 3-string vertex} }
\end{picture} \\
\mbox{(a)} & \mbox{(b)}
\end{array}$
\caption{(a) An open string in ground state with momentum $k$.
(b) The 3-string vertex.}
\end{center}
\end{figure}

In general, by patching several 3-string vertices together, 
we can get an $n$-string vertex, 
which will be represented by an $n$-polygon. 
The scattering amplitude for this $n$-string tree level diagram $(n \geq 3)$ 
is given by a phase proportional to the area of the polygon.
Due to the conservation of area, 
this agrees with the product of the phase factors for the $(n-2)$ 3-string vertices
composing the $n$-string vertex.

\section{Generalization to nonassociative space} \label{NAS}

In this section we generalize the heuristic derivation of noncommutative space 
from open strings in $B$-field background to 
the derivation of nonassociative space from open membranes in $C$-field background. 
%%% 0108 %%%
In M theory, let us consider open membranes with their boundaries moving along 
a flat M5-brane extended in the directions $(123456)$
(see for example, \cite{Seiberg:1999vs, Sakaguchi:2006ph}, for the
constraint on $C$-field background). 
We turn on a constant $C$-field background on the M5-brane with 
the only non-vanishing components $C_{123}=C, C_{456}$, 
which allows us to think of the 6-dimensional world-volume of M5-brane 
as a product space $\mathbf{R}^3 \times \mathbf{R}^3$. 
We will only focus on the directions $(123)$. 
A complete description including directions $(456)$ is straightforward.
%% p %%
\footnote{Strictly speaking, one of the 6 directions of an M5-brane is time-like. 
Assuming that the time direction is $X^6$,
$C_{456}$ is bounded from above by a critical value, 
similar to the electric flux on a D-brane.
This issue is not the focus of this paper.}
%% 0108 %%

For a large constant $C$-field background in the $X_1$-$X_2$-$X_3$ direction, 
the world-volume action is dominated by the term
\be \label{S}
S = \int C \; dX^1 dX^2 dX^3 = \int d\tau d^2 \sigma \; C \; \eps_{ijk} \; \del_{\tau} X^i \del_{\s_1} X^j \del_{\s_2} X^k, 
\ee
where we ignored spacetime coordinates in other directions. 
Since the integrand is a total derivative,
the action (\ref{S}) can also be interpreted as the action for a closed string
in a large $H$-field background with
\be
H = dB, \qquad B = C \eps_{ijk} X^i dX^j dX^k.
\ee
What we will call open membranes can also be called closed strings.

According to this action, the canonical momentum is 
\be \label{PCXX}
P_i = C\; \eps_{ijk} \int d X^j d X^k. 
\ee
Therefore, the momentum is perpendicular to the 
the world volume of the open membrane, 
and the magnitude of the momentum equals $C$ times the area of the membrane. 
The conservation of momentum is geometrically represented by the conservation of area (as a 3-vector).

The expression (\ref{PCXX}) agrees with analysis 
for the BFSS matrix model.
It was shown in \cite{Chu:1999ne,Seiberg:1999vs} that the effect of 
turning on a $C$-field background is equivalent to a shift of 
the momentum 
\be
P_i \rightarrow P_i -\frac{i}{2} C_{ijk} [X_i, X_j],
\ee
where $P_i, X_j, X_k$ are matrix variables. 

In analogy with the derivation of the uncertainty relation (\ref{UR}) 
for the noncommutative space, 
(\ref{PCXX}) suggests a generalized uncertainty relation. 
Consider an open membrane moving in the $X_1$- direction. 
Eq. (\ref{PCXX}) implies  
\be
\d P_1 \sim C \d X_2 \d X_3 . 
\ee
Together with the usual uncertainty relation (\ref{usualUR}), 
it leads to the generalized uncertainty relation 
\be \label{UR3}
\d X_1 \d X_2 \d X_3 \geq \frac{1}{C}. 
\ee
Here and below we will take the convention that $C > 0$.

While the uncertainty relation (\ref{UR}) can be viewed as a reflection of 
the noncommutative nature of space, 
the uncertainty relation (\ref{UR3}) can be viewed as a reflection 
of the nonassociative nature of space.
%%% 0108 %%%
Unlike the noncommutative nature of spacetime, 
for which spacetime coordinates come in conjugate pairs, 
spacetime coordinates are divided into groups of 3 by nonassociativity.
%% 0108 %%

Consider the tree level diagram of 4 external legs. 
Each leg is an open membrane with given momentum, 
which determines the area of the world-volume cross section. 
But the shape of the cross section is not fixed. 
Following the same steps in the previous section, 
we assume that the scattering amplitude is given by the configuration 
which minimizes the world-volume.
Apparently, the unique minimal volume configuration should have
the 4 legs attached to the 4 faces of a tetrahedron.
(See Fig. \ref{figmembrane} (b).) 
As a result, the shape of the cross section of each external leg must be a triangle, 
and each triangle has an area equal to $C$ times the momentum. 

\begin{figure}[ptb]\label{figmembrane}
\setlength{\unitlength}{3pt}
\par
\begin{center}
$\begin{array}{cc}
\begin{picture}(40,40)(-20,-20)
\put(8,24){\vector(-1,-3){8}}
\put(0,12){$\vec \ell_3$}
\put(0,0){\vector(3,-1){24}}
\put(10,-7){$\vec \ell_1$}
\put(24,-8){\vector(-1,2){16}}
\put(20,2){$\vec \ell_2$}
\put(0,0){\line(-1,0){35}}
\put(8,24){\line(-1,0){35}}
\put(24,-8){\line(-1,0){35}}
\put(12, 7){\vector(1,0){10}}
\put(17,8){$\vec k$}
\end{picture}
&
\begin{picture}(80,80)(-40,-40)
% draw tetrahedron
\put(16,24){\vector(-2,-3){16}}
\put(4,12){$\vec \ell_1$}
\put(16,24){\vector(1,-2){12}}
\put(17, 5){$\vec \ell_2$}
\put(16,24){\vector(1,-4){8}}
\put(24, 10){$\vec \ell_3$}
\put(0,0){\line(1,0){28}}
\put(0,0){\line(3,-1){24}}
\put(28,0){\line(-1,-2){4}}

\put(16.2,24){\vector(-2,-3){16}}
\put(16.2,24){\vector(1,-2){12}}
\put(16.2,24){\vector(1,-4){8}}
\put(0.2,0){\line(1,0){28}}
\put(0,0.2){\line(3,-1){24}}
\put(28.2,0){\line(-1,-2){4}}

% open membrane #1
{\color{Dark}
\put(0,0){\line(-4,1){40}}
\put(16,24){\line(-4,1){40}}
\put(24,-8){\line(-4,1){40}}
\put(-40,10){\line(2,3){16}}
\put(-40,10){\line(3,-1){24}}
\put(-24,34){\line(1,-4){8}}
% open membrane #2
\put(0,0){\line(-1,6){6}}
\put(16,24){\line(-1,6){6}}
\put(28,0){\line(-1,6){6}}
\put(-6,36){\line(1,0){28}}
\put(-6,36){\line(2,3){16}}
\put(10, 60){\line(1,-2){12}}
% open membrane #3
\put(24,-8){\line(4,1){40}}
\put(16,24){\line(4,1){40}}
\put(28,0){\line(4,1){40}}
\put(64,2){\line(-1,4){8}}
\put(64,2){\line(1,2){4}}
\put(68,10){\line(-1,2){12}}
% open membrane #4
\put(24,-8){\line(-1,-6){5}}
\put(0,0){\line(-1,-6){5}}
\put(28,0){\line(-1,-6){5}}
\put(-5,-30){\line(1,0){28}}
\put(-5,-30){\line(3,-1){24}}
\put(19,-38){\line(1,2){4}}}
\end{picture} \\
\mbox{(a)} & \mbox{(b)}
\end{array}$
\caption{(a) The open membrane or closed string. 
(b) The 4-membrane vertex} 
\end{center}
\end{figure}

Denoting the edges of the tetrahedron in Fig. \ref{figmembrane} (b) by $\ell_i$'s as shown,
the volume of the tetrahedron is
\be
\frac{1}{6}\vec \ell_1\cdot(\vec \ell_2\times \vec \ell_3)
\ee
and its contribution to the action is
%% p %%
\footnote{Here we assume that the saddle point approximation still works. 
For simplicity, we also dropped a constant factor.}
\be \label{S0}
S = C \vec \ell_1\cdot(\vec \ell_2\times\vec \ell_3).
\ee
%which can be either positive or negative depending on the orientation of the tetrahedron.

For given external momenta $k_a$ ($a = 1, 2, 3, 4$), 
we have
\be
\vec k_1 = C \vec \ell_1 \times \vec \ell_2, \quad \vec k_2 = C \vec \ell_2 \times \vec \ell_3, 
\quad \vec k_3 = C \vec \ell_3\times \vec \ell_1.
\ee
It is then straightforward to check that
\be
\vec k_1 \cdot (\vec k_2 \times \vec k_3) = C^3 \left[ \vec \ell_1 \cdot (\vec \ell_2 \times \vec \ell_3) \right]^2.
\ee
The minimal volume of the tetrahedron is thus
\be
\frac{1}{6}C^{-3/2} \sqrt{| \vec k_1 \cdot (\vec k_2 \times \vec k_3) |}. 
\ee
Due to energy-momentum conservation, one can choose any 3 of the 4 momenta $k_a$
to compute the volume.
%%% 0108 %%%
The conservation of energy-momentum is guaranteed by the fact that the tetrahedron is a closed surface
\be
\vec k_1 + \vec k_2 + \vec k_3 + \vec k_4 = C \left(
\vec \ell_1 \times \vec \ell_2 + \vec \ell_2 \times \vec \ell_3 + \vec \ell_3 \times \vec \ell_1 
+ (\vec \ell_3 - \vec \ell_1) \times (\vec \ell_2 - \vec \ell_1) \right) = 0.
\ee
%% 0108 %%

The phase factor of the 4-point interaction vertex is thus
\be
e^{iC\vec \ell_1 \cdot (\vec \ell_2 \times \vec \ell_3)} = e^{\pm iC^{-1/2}\sqrt{|\vec k_1 \cdot (\vec k_2 \times \vec k_3)|}}. 
\ee
The sign of the exponent is determined by the orientation of the tetrahedron, 
which is an information encoded in $(\vec \ell_1, \vec \ell_2, \vec \ell_3)$, 
but not in $(\vec k_1, \vec k_2, \vec k_3)$. 
For each tetrahedron spanned by the 3 vectors $(\vec \ell_1, \vec \ell_2, \vec \ell_3)$, 
there is another tetrahedron obtained by a spatial inversion $(-\vec \ell_1, -\vec \ell_2, -\vec \ell_3)$ 
corresponding to the same external momenta $(\vec k_1, \vec k_2, \vec k_3)$, 
but with the opposite orientation, 
so that the exponent of the phase factor (\ref{S0}) gets an opposite sign. 
Drawing the tetrahedron corresponding to $(-\vec \ell_1, -\vec \ell_2, -\vec \ell_3)$, 
one can see that it has a ``negative volume'' compared with figure \ref{figmembrane} (b) 
in the sense that the tetrahedron is the overlap of the (fat) external legs, 
while in figure \ref{figmembrane} (b) the external legs do not overlap. 

In the above we have assumed that the only quantum number needed 
to label the ground states of the open membrane is the momentum. 
Both orientations $(\vec \ell_1, \vec \ell_2, \vec \ell_3)$ and $(-\vec \ell_1, -\vec \ell_2, -\vec \ell_3)$ 
are allowed and should be summed over.
The phase factors add up for ground state interactions to give
\be
\cos\left(C \vec \ell_1\cdot(\vec \ell_2\times\vec \ell_3)\right) = 
\cos\left(C^{-1/2}\sqrt{|\vec k_1 \cdot (\vec k_2 \times \vec k_3)|}\right).
\ee
For large $C$, this phase factor differs from the trivial case ($C = \infty$) by the correction 
\be
\cos\left(C^{-1/2}\sqrt{|\vec k_1 \cdot (\vec k_2 \times \vec k_3)|}\right) - 1
\simeq \frac{1}{2C}|\vec k_1 \cdot (\vec k_2 \times \vec k_3)|.
\ee
It is interesting to note that this factor is proportional to the prefactor 
due to the classical Nambu bracket 
\be \label{CNB}
\{ \psi_1, \psi_2, \psi_3 \} \equiv \eps^{ijk}\del_i \psi_1 \del_j \psi_2 \del_k \psi_3 
\ee
for plane waves $\psi_i = \exp(i\vec k_i \cdot \vec x)$.
One may thus imagine that the interaction due to the $C$-field background 
corresponds to a quantum version of the Nambu bracket in the effective field theory. 

It is also interesting to notice that the open membrane naturally takes 
the shape of a triangle. 
This is analogous to the fact that we only need the two endpoints 
of the open string to describe the noncommutativity of space. 
We will see that only 3 points on the open membrane is sufficient 
to derive the essential features of nonassociative space.

%% m %%
This feature is, of course, limited to the four-point function.
When we have more external lines, in order to minimize the
volume, we have to include polygons in general.
In this paper, instead of including these generalized membrane
field, we first limit ourselves to consider a toy model
by using the triangles 
as a good starting point to a regularized membrane theory.
One reasoning toward this simplification is that the lattice
gravity in three dimension is based on the tetrahedra.
We will also discuss later a generalization to include the membrane
with the polygon shape.

\section{Membrane field on a triangle and products}

Motivated from preceding arguments,
we introduce a truncated version of the membrane field
which captures the basic nature of the 
membrane fields in a large $C$-field background.

\paragraph{Definition of membrane field truncated on a triangle}
As we have seen, the four membrane interaction 
is described by a tetrahedron where membranes
are attached to four triangles.
So we start from the truncated open membrane 
with the shape of a triangle, 
determined by the position vectors 
$\{\vec x_i\}_{i=1}^3$ of its 3 corners. 
We write the truncated membrane wave function as
$\Phi(\vec x_1, \vec x_2, \vec x_3)$.
The relation between $\vec x_i$ and $\vl_i$ in the previous section is
$\vl_i=\vec x_{i+1}-\vec x_i$ ($i=1,2,3$, $\vl_4\equiv \vl_1)$.

Since a triangle is specified by the positions of the corners
but not their order, we impose the cyclicity of the membrane field,
\be
\Phi(\vec x_1, \vec x_2, \vec x_3)=
\Phi(\vec x_2, \vec x_3, \vec x_1)=
\Phi(\vec x_3, \vec x_1, \vec x_2)\,.
\ee
On the other hand, for the triangle with the opposite orientation,
we need to impose
\be
\Phi(\vec x_3, \vec x_2, \vec x_1)=
\Phi(\vec x_2, \vec x_1, \vec x_3)=
\Phi(\vec x_1, \vec x_3, \vec x_2)=
(\Phi(\vec x_1, \vec x_2, \vec x_3))^*\,.
\ee
We will refer to these properties as the Hermicity of the membrane fields.

To impose the constraint (or equation of motion) that the
mementum is proportional to the area of the triangle, 
we use the momentum representation.
We define a change of variable from $\vec x_1, \vec x_2, \vec x_3$
to $\vec x_c=\frac{1}{3}(\vec x_1+\vec x_2+\vec x_3)$
and $\vl_1, \vl_2, \vl_3$, \footnote{
$\vl_3$ is not an independent variable since $\vl_3=-\vl_1-\vl_2$.
We keep it, however, to make the formulae in later
discussion more symmetric.
}
\begin{equation}
\Phi(\vec x_1, \vec x_2, \vec x_3) = 
\bar{\Phi}(\vec x_c; \vec \ell_1, \vec \ell_2, \vl_3),
\end{equation}
and carry out the Fourier transformation,
with respect to the center of mass coordinate $\vec x_c$,
\begin{equation}
\bar{\Phi}(\vec x_c; \vec \ell_1, \vec \ell_2,\vl_3) = \int d^3 \vec k
e^{i\vec k\cdot \vec x_c} \tilde{\Phi}(\vec k; \vec \ell_1, \vec \ell_2,\vl_3).
\end{equation}
The constraint can now be written in terms of the Fourier image as
% The ground states of the open membrane 
% correspond to functions of the form
\be\label{onshell}
(\vec k - C \vec \ell_1 \times \vec \ell_2) 
\hat{\Phi}(\vec k; \vec \ell_1, \vec \ell_2,\vl_3) = 0\,.
\ee
It can be solved as
\begin{equation}
\tilde{\Phi}(\vec k; \vec \ell_1, \vec \ell_2,\vl_3)
= \delta^{(3)}(\vec k - C \vec \ell_1\times \vec \ell_2)
{\hat \Phi}(\vec \ell_1, \vec \ell_2,\vl_3)\,.
\end{equation}
In terms of the original variables,
\be\label{Phi1}
\Phi(\vec x_1, \vec x_2, \vec x_3) = 
\bar{\Phi}(\vec x_c; \vec \ell_1, \vec \ell_2,\vl_3)
=e^{i C\vec x_c\cdot (\vl_1\times \vl_2)}
{\hat \Phi}(\vec \ell_1, \vec \ell_2,\vl_3)\,.
\ee

The coefficient $\hat\Phi(\vec \ell_1, \vec \ell_2,\vl_3)$
which depends on the shape of the triangle carries the
information of the membrane field.  We can expand it as
\be\label{fiex}
\hat\Phi(\vec \ell_1, \vec \ell_2,\vl_3)
=A\phi(C \vl_1\times \vl_2)+\cdots\,.
\ee
The first component $\phi(\vec k)=\phi(C \vl_1\times \vl_2)$
depends only on the momentum of the membrane field.
It describes the ground state of the membrane.
We will refer to it as the tachyon field by using 
the terminology of (bosonic) string field theory.
$A$ is a normalization constant which will be determined in a moment.
On the other hand, the subsequent terms  ($\cdots$) describe
the excited states of the membrane.
In the construction of the field theory in Sec.\ref{sec:field_theory},
we will focus on the field theory of the ground state $\phi(\vec k)$.
The Hermicity of the membrane in terms of $\hat \Phi$ is
written as $\hat \Phi(\vl_{\sigma(1)},\vl_{\sigma(2)},\vl_{\sigma(3)})
=\hat \Phi(\vl_1,\vl_2,\vl_3)$ for even permutations $\sigma\in S_3$
and $\hat \Phi(\vl_{\sigma(1)},\vl_{\sigma(2)},\vl_{\sigma(3)})=\hat \Phi(\vl_1,\vl_2,\vl_3)^*$ for odd permutations.
For the tachyon field, it is simply the condition for a real scalar field
$\phi(-\vec k)=\phi(\vec k)^*$.

The inner product between membrane fields 
is most naturally defined by
the overlap condition
\bea\label{naive_inner}
\int d\vec x_1d\vec x_2 d \vec x_3
\Phi_1(\vec x_3, \vec x_2, \vec x_1)
\Phi_2(\vec x_1, \vec x_2, \vec x_3)
=\int d\vec x_1d\vec x_2 d \vec x_3
\Phi_1(\vec x_1, \vec x_2, \vec x_3)^*
\Phi_2(\vec x_1, \vec x_2, \vec x_3)\,.
\eea
If we use the on-shell condition (\ref{onshell}), however,
it contains an infinite constant since
the integration over $\vec x_c$ produces
$\delta^{(3)}(\vec k-\vec k)=\delta^{(3)}(0)$.
So we define the inner product after getting rid of
this factor
\bea
(\Phi_1,\Phi_2)&=&\int d\vl_1d\vl_2\vl_3
\hat\Phi_1(\vl_1,\vl_2, \vl_3)^*
\hat \Phi_2(\vl_1,\vl_2, \vl_3)\delta^{(3)}(\vl_1+\vl_2+\vl_3)\,.
\eea
The normalization constant in (\ref{fiex}) is determined to have
a standard norm between the ground state wave function
$\phi$,
$
(\Phi_1,\Phi_2)=\int d\vec k \phi_1(-\vec k)\phi_2(\vec k)
=\int d\vec k \phi_1(\vec k)^*\phi_2(\vec k)
$. Namely, 
\be
A^{-2}=\int d\vl_1 d\vl_2 \delta^{(3)}(\vec k-C\vl_1\times\vl_2)\propto |\vec k|^{-1}\,.
\ee

%%%%%%%%%%%%%%%%%%%%%%%%%%%%%%%%%%%%%%%%%%%%%%
\paragraph{Cubic and quartic product for membrane field}

\begin{figure}[bpt]
\begin{center}
\includegraphics[scale=0.8]{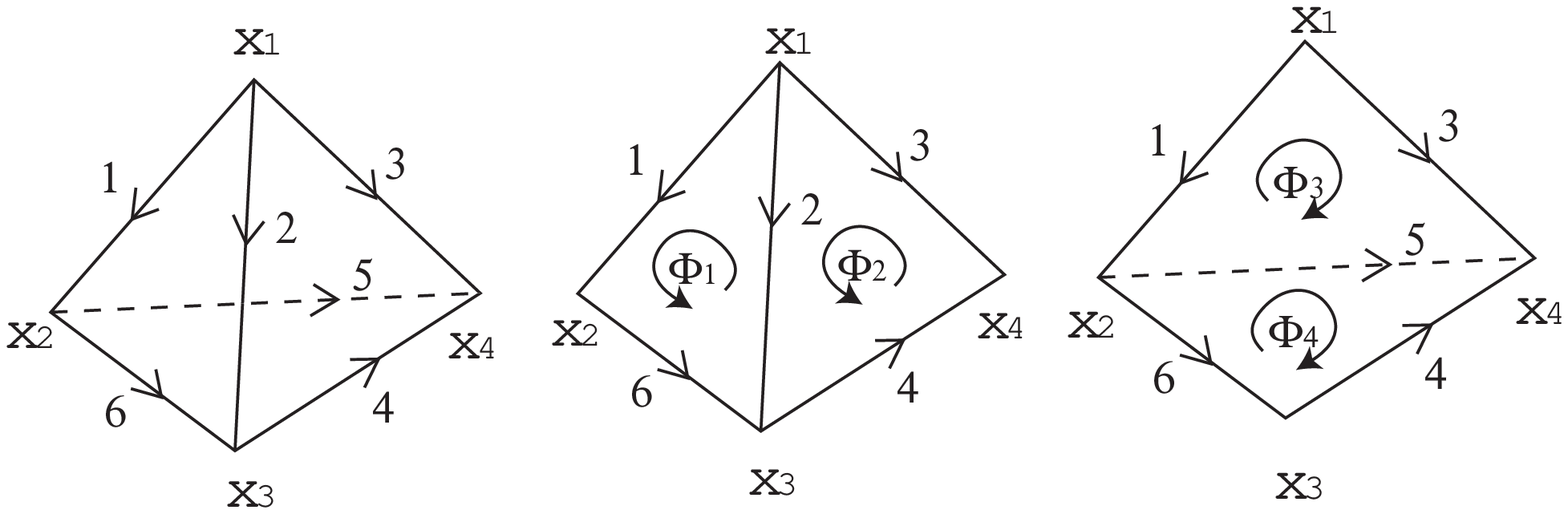}
\end{center}
\caption{Patching of the faces at the tetrahedron.
The vector symbol
$\vl$ is omitted at the edges and only their label is denoted.
Curly arrow shows the orientation
of each faces.  The sign of the
vector $\vl_i$ in $\Phi$  is determined by whether
the arrow at the edge is the same (resp. opposite)
direction to the curly arrow of the face which represent
$\Phi$. }
\label{fig:tetrahedra}
\end{figure}

The simplest interaction for the membrane fields on a triangle
is the quartic product as we have seen in the previous section.
We define it through the overlap condition for the boundaries
of the triangle as in the inner product,
\be\label{qprod}
\int d^3 \vec x_1 d^3 \vec x_2 d^3 \vec x_3 d^3 \vec x_4
\Phi_1(\vec x_1, \vec x_2, \vec x_3) \Phi_2(\vec x_1, \vec x_3, \vec x_4) 
\Phi_3(\vec x_1, \vec x_4, \vec x_2) \Phi_4(\vec x_2, \vec x_4, \vec x_3).
\ee
We illustrate the locations of the corners in 
Fig.\ref{fig:tetrahedra}.  
We need to be careful in the orientation
of each triangle to specify the order of $\vec x_i$ in each
membrane field.
It may be regarded as a discretized version of the interaction 
for the closed string which describes the boundary of the flat
membrane. 
%%% 0108 %%%
% Sorry but I don't understand the following sentence
% It might sound strange that we
% have the interaction term of string field theory to describe
% the membrane. It is possible since we used the approximation that
% the membrane is flat and its dynamics is encoded in the boundary
% degree of freedom.
%% 0108 %%

%%%%%%%%%%%%%%%%%%%%%%%%%%%%%%%%%%%%%%%%%%%%%%%%%%%%%
%%%%%%%%%%%%%%%%%%%%%%%%%%%%%%%%%%%%%%%%%%%%%%%%%%%%%
We put the membrane field (\ref{Phi1}) into eq.(\ref{qprod}),
the phase factor before the integration becomes
(the assignment of $\vl_i $ is written in Fig.\ref{fig:tetrahedra}),
\bea
&&iC( \vec x^1_c\cdot(\vl_1\times \vl_2)
+\vec x^2_c\cdot(\vl_2\times \vl_3)
+\vec x^3_c \cdot(\vl_3\times \vl_1)
+\vec x^4_c\cdot(\vl_4\times \vl_5))\nonumber\\
&&=  iC( \vec y^1_c\cdot(\vl_1\times \vl_2)
+\vec y^2_c\cdot(\vl_2\times \vl_3)
+\vec y^3_c \cdot(\vl_3\times \vl_1)
+\vec y^4_c\cdot(\vl_4\times \vl_5))\nonumber\\
&& ~~~~~~~~~+iCx^t_c (\vl_1\times \vl_2+\vl_2\times \vl_3
+ \vl_3\times \vl_1 +\vl_4\times \vl_5)\,,
\eea
where $\vec x^t_c=\frac{\vec x_1+\vec x_2+\vec x_3+\vec x_4}{4}$
and $\vec y^i_c= \vec x^i_c-\vec x^t_c$.
The expression in the third line vanishes because of the
momentum conservation at the vertex.  On the other hand,
the expression in the second line becomes 
\be
iC\vl_1\cdot(\vl_2\times \vl_3)= iC V
\ee
where $V$ is the volume of the tetrahedron.  This is
the phase factor which we observed in the previous section.
Since the dependence on the center of the mass coordinate $\vec x^t_c$
vanishes, the integration on this variable gives rise to $\delta^{(3)}(0)$ again.
So it will be more useful to define the quartic product 
$\left<\Phi,\Phi,\Phi,\Phi\right>$ by
the integration over the relative coordinates $\vl_i$,
\begin{eqnarray}
 \left<\Phi_1,\Phi_2,\Phi_3,\Phi_4\right>
&=& \int d\vec \ell_1 \cdots d\vec\ell_6 
\,e^{iC\vec\ell_1\cdot(\vec\ell_2\times\vec\ell_3)}\,\cdot
\nonumber\\
&&\cdot\hat\Phi_1(\vec \ell_1,\vec \ell_6, -\vec \ell_2)
\hat\Phi_2(\vec \ell_2,\vec \ell_4, -\vec \ell_3)
\hat\Phi_3(-\vec \ell_1,\vec \ell_3, \vec \ell_5)
\hat\Phi_4(-\vec \ell_4,-\vec \ell_6, -\vec \ell_5)\cdot\nonumber\\
&&\cdot \,
\delta^{(3)}(\vec \ell_1+\vec \ell_6,-\vec \ell_2)
\delta^{(3)}(\vec \ell_2+\vec \ell_4 -\vec \ell_3)
\delta^{(3)}(-\vec \ell_1+\vec \ell_3+ \vec \ell_5)\,.
\label{quartic}
\end{eqnarray}
In the next section, we introduce the interaction
term for the tachyon field by using this product.
As in the inner product, we need to be careful in choosing
the normalization.  We postpone this issue to the next
section since it is more complicated.

Since the quartic product 
defines a map ${\cal H}^{\otimes 4}\rightarrow \mathbf{C}$,
we can obtain a triplet product $[\Phi,\Phi,\Phi]$
${\cal H}^{\otimes 3}\rightarrow \mathcal{H}$
by conjugation with respect
to one membrane field, namely by requiring
\be
\left<\Phi_1,\Phi_2,\Phi_3,\Phi_4\right>=([\Phi_1,\Phi_2,\Phi_3],\Phi_4).
\ee
In terms of the $\vec x$ representation,
\begin{equation} \label{tpm}
[\Phi_1,\Phi_2,\Phi_3](\vec x_1, \vec x_2, \vec x_3) = \int d^3 \vec x\,
\Phi_1(\vec x, \vec x_2, \vec x_1) \Phi_2(\vec x, \vec x_3, \vec x_2) 
\Phi_3(\vec x, \vec x_1, \vec x_3).
\end{equation}
This is the natural definition of the triplet product for truncated open membranes. 
It is obviously different from the triplet product we defined for the ground states. 
But we will see below that, 
restricting the truncated open membrane to ground states in the $C$-field background, 
this triplet product (\ref{tpm}) reduces to exactly the same product (\ref{cubicproduct}) for ground states.

The Fourier transform
of the product $[\Phi_1,\Phi_2,\Phi_3]$ is given by
\begin{eqnarray}
\widetilde{[\Phi_1,\Phi_2,\Phi_3]}(\vec k; \vec \ell_1, \vec \ell_2,-\vl_1-\vl_2)
= \int d^3 \vec k_1 d^3 \vec k_3 d^3 \vec \ell \;
e^{\frac{i}{3}(\vec k\cdot \vec \ell + \vec k_1 \cdot \vec \ell_2 - \vec k_3 \cdot \vec \ell_1)}
{\tilde\Phi_1}(\vec k_1; -\vec \ell, \vec \ell-\vec \ell_1,\vl_1) \nonumber \\
{\tilde\Phi_2}(\vec k-\vec k_1-\vec k_3; 
\vec \ell_1-\vec \ell, \vec \ell+\vec \ell_2,-\vl_1-\vl_2)
{\tilde\Phi_3}(\vec k_3; -\vec \ell-\vec \ell_2, \vec \ell,\vl_2).
\end{eqnarray}
{}From this, we can write the on-shell part of the triplet product
as
\bea
&&\widetilde{[\Phi_1,\Phi_2,\Phi_3]}(\vec k; \vec \ell_1, \vec \ell_2,-\vl_1-\vl_2)
 \equiv
\delta^{(3)}(\vec k-C \vec \ell_1\times\vec \ell_2)
\widehat{[\Phi_1,\Phi_2,\Phi_3]}
(\vec \ell_1, \vec \ell_2, -\vec \ell_1-\vec \ell_2),\\
&&\widehat{\left[\Phi_1,\Phi_2,\Phi_3\right]}
(\vec \ell_1, \vec \ell_2,-\vl_1-\vl_2)\nonumber\\
&& ~~~~~ =
 \int d\vec \ell
\,e^{iC\vec\ell\cdot(\vec\ell_1\times\vec\ell_2)}
\hat\Phi_1(-\vec \ell, \vec \ell-\vec \ell_1,\vl_1)
\hat\Phi_2(\vec \ell_1-\vec \ell, \vec \ell+\vec \ell_2,-\vl_1-\vl_2)
\hat\Phi_3(-\vec \ell-\vec \ell_2, \vec \ell,\vl_2)\,.
\label{cubicproduct}
\eea

%%%%%%%%%%%%%%%%%%%%%%%%%%%%%%%%%%%%%%%%%%%%%%%
\section{Scalar field theory on nonassociative space}
\label{sec:field_theory}

Let us consider an analogy of the scalar field theory
for the membrane field.
The dimensions of the spacetime should be taken as a multiple of $3$. 
We will just consider the case of 3-dimensions here for simplicity. 
While it is possible to include the complicated interaction terms 
which can be constructed by using various product formula
in Sec. \ref{sec:higher} below, we first examine
the simplest action for the membrane field 
$\Phi(\vec k)$
\begin{eqnarray}
S&=&S_{kin}-
S_{int}\,,
%\frac{g}{4!}[\Phi,\Phi,\Phi,\Phi]^{(4)}\,,
\end{eqnarray}
where the first term is the ordinary kinetic term
\be
S_{kin} = \int d^3 \vec k \Psi^{\ast}(\vec k) (\vec k^2 + m^2) \Psi(\vec k),
\ee
and the second term is the 4-point interaction vertex obtained in the previous section
\be
S_{int} = \frac{g}{4!} \int d^3 \vec k_1 d^3 \vec k_2 d^3 \vec k_3 \cos\left(C^{-1/2}\sqrt{|\vec k_1 \cdot (\vec k_2 \times \vec k_3)|}\right) 
\Psi(\vec k_1)\Psi(\vec k_2)\Psi(\vec k_3)\Psi(-\vec k_1-\vec k_2-\vec k_3).
\ee

One may feel a bit unease with the absolute value involved in the definition of $S_{int}$, 
and in view of the notion of truncated open membrane which we introduced above, 
it is worthwhile to construct an equivalent description of the field theory 
in terms of the wave functions $\Phi(\vec \ell_1, \vec \ell_2, \vec \ell_3)$ labeled by $\vec \ell$'s. 
Since the only quantum number of the open membrane ground state is the momentum,
if we denote the ground state with momentum $\vec k$ by $|\vec k\rangle$, 
a generic wave function for the ground state is 
\be
|\Phi\rangle = \int d^3\vec k \Psi(\vec k) |\vec k\rangle. 
\ee
On the other hand, in terms of the wave function $\Phi(\vec \ell_1, \vec \ell_2, \vec \ell_3)$, it should be
\be
|\Phi\rangle = \int d^3 \vec \ell_1 d^3 \vec \ell_2 \Phi(\vec \ell_1, \vec \ell_2, -\vec \ell_1-\vec \ell_2) 
| C \vec \ell_1\times \vec \ell_2\rangle.
\ee
The connection between these two representations is therefore
\be
\Psi(\vec k) = \int d^3 \vec \ell_1 d^3 \vec \ell_2 \delta^{(3)}(\vec k - C \vec \ell_1 \times \vec \ell_2) 
\Phi(\vec \ell_1, \vec \ell_2, -\vec \ell_1-\vec \ell_2).
\ee 

In terms of $\Phi(\vec \ell_1, \vec \ell_2, \vec \ell_3)$, 
the kinetic term of the action is
\bea 
S_{kin}&=&\frac{1}{2}
\int d^{3}\vec \ell_1 d^{3}\vec \ell_2 d^{3}\vec \ell'_1 d^{3}\vec \ell'_2 \;
\Phi^{\ast}(\vec \ell_1, \vec \ell_2, -\vec \ell_1-\vec \ell_2)
\left(C^2(\vec \ell_1\times \vec \ell_2)^2+m^2\right)\cdot\nonumber\\
&&~~~~~\cdot\Phi(\vec \ell'_1, \vec \ell'_2, -\vec \ell'_1-\vec \ell'_2)\,
\delta^{(3)}(\vec \ell_1 \times \vec \ell_2 - \vec \ell'_1 \times \vec \ell'_2)
\,.\label{Skin2}
\eea
The interaction term is 
\bea
S_{int} &=& \int d^3\vec \ell_1 d^3\vec \ell_2 d^3\vec\ell_3 \,
{\cal N}(\vec \ell)\,e^{iC\vec\ell_1\cdot(\vec\ell_2\times\vec\ell_3)}\,
\Phi(\vec \ell_1-\vec \ell_2,-\vec \ell_1, \vec \ell_2)
\Phi(-\vec \ell_2,\vec \ell_3, \vec \ell_2 - \vec \ell_3)
\,\cdot \nonumber\\
&&\Phi(-\vec \ell_3,\vec \ell_1, \vec \ell_3-\vec \ell_1)
\Phi(\vec \ell_3-\vec \ell_2,\vec \ell_1-\vec \ell_3, \vec \ell_2-\vec \ell_1)\, ,
\eea
where $\vec \ell_1, \vec \ell_2, \vec \ell_3$ spans a tetrahedron 
whose faces match with the arguments of the 4 $\Phi$'s.
The normalization factor ${\cal N}(\vec \ell)$ will be determined later.

The kinetic term (\ref{Skin2}) looks a bit complicated. 
One may wonder what happens if we make a more naive choice
\be
S'_{kin}(\Phi)=\frac{1}{2}
\int d^{9}\vec \ell \;
\delta^{(3)}(\vl_1+\vl_2+\vl_3)
\Phi^{\ast}(\vl_1,\vl_2,\vl_3)
\left(C^2(\vec\ell_1\times\vec\ell_2)^2+m^2\right)
\Phi(\vl_1,\vl_2,\vl_3)\,.
\ee
It implies a propagator of the form
\be
\langle \Phi(\vl')\Phi(\vl)\rangle' =
\frac{1}{C^2(\vec \ell_1\times \vec \ell_2)^2+m^2}
\delta^{(3)}(\vl+\vl')\,.
\ee
However, this definition leads to a Feynman rule which is very restrictive. 
For example, the internal loop momentum of the one-loop diagram 
in figure \ref{OneLoopTwoLoop} (b) is completely fixed by the external legs. 
Furthermore, the external legs on the left must have 
exactly the same labels $\vec \ell$ as the other two external legs on the right. 
This is a result of the fact that the labels $\vec \ell$ on any two faces of a tetrahedron 
completely fix the labels on the other two faces, 
and the fact that this propagator does not connect cross sections with different shapes.

%%%%%%%%%%%%%%%%%%%%%%%%%%%%%%%%%%%%%%%%%%%%
\subsection{Feynman diagrams} 

In this section we consider basic Feynman diagrams 
involving the 4-membrane interaction vertex $[\Phi,\Phi,\Phi,\Phi]^{(4)}$. 

Before we start computing Feynman diagrams, 
we need to understand the phase space. 
For particles with no internal degrees of freedom, 
the state of a free particle can be specified by its momentum $p$.
For the triangular open membrane, 
its internal degrees of freedom are parameterized by two 3-vectors 
($\vec \ell_1$, $\vec \ell_2$) constrained by 
\be
\vec p = C \vec \ell_1 \times \vec \ell_2
\ee
for the state with momentum $\vec p$.
Hence the phase space for the internal degrees of freedom 
has the measure 
\be \label{measure}
d^3 \vec \ell_1 \; d^3 \vec \ell_2 \; \delta^{(3)}(\vec p - C \vec \ell_1 \times \vec \ell_2).
\ee
As $\vec p$ and $\vec \ell_1$ are always perpendicular to each other, 
we can use $\vec p$, $\vec \ell_1$ and $\vec p \times \vec \ell_1$ 
as an orthogonal basis of vectors and expand $\vec \ell_2$ as
\be
\vec \ell_2 = a \vec p \times \vec \ell_1 + b \vec \ell_1 + c \vec p.
\ee
Then 
\be
d^3 \vec \ell_2 = da \; db \; dc \; (\vec p \times \vec \ell_1)^2
\ee
and the measure (\ref{measure}) can be simplified as
\be
d^3 \vec \ell_1 \; db \; \frac{1}{C^2} \delta(\vec p \cdot \vec \ell_1),
\ee
while $\vec \ell_2$ is now fixed to be
\be
\vec \ell_2 = \frac{1}{C \ell_1^2}(\vec p \times \vec \ell_1) + b \vec \ell_1.
\ee

%%%%%%%%%%%%%%%%%%%%%%%%%%%%%%%%%%%%%%%%%%%%
\subsubsection{Tree diagram} 

Let us consider the tree level diagram of 4 external legs with a single interaction vertex.
For the 4 membranes to fit onto the 4 edges of a tetrahedron defined by 
$(\vec \ell_1, \vec \ell_2, \vec \ell_3)$, 
the external legs must have triangular cross sections given by 
$(\vec \ell_1, \vec \ell_2)$, $(\vec \ell_2, \vec \ell_3)$, 
$(\vec \ell_3, \vec \ell_1)$ and $(\vec \ell_1-\vec \ell_2, \vec \ell_3-\vec \ell_2)$. 

For given momenta of the external legs $\vec p_i$ $(i = 1,2,3,4)$,
the tree level diagram which is simply the 4-membrane vertex 
is supposed to be given by 
\be \label{4v}
\cos\left(C^{-1/2} \sqrt{|\vec p_1 \cdot (\vec p_2 \times \vec p_3)|}\right),
\ee
where the sign is determined by the orientation of the tetrahedron.
We would like to see how this can be derived from 
integration over the phase space of internal degrees of freedom.
The 4-membrane vertex is
\bea
{\cal A}_0 &=& \int d^3 \vec \ell_1 d^3 \vec \ell_2 d^3 \vec \ell_3 \, {\cal N}(\vec \ell) \,
\delta^{(3)}(\vec p_1 - C \vec \ell_1 \times \vec \ell_2)\;
\delta^{(3)}(\vec p_2 - C \vec \ell_2 \times \vec \ell_3)\;
\delta^{(3)}(\vec p_3 - C \vec \ell_3 \times \vec \ell_1)\cdot \nn \\
&&\delta^{(3)}\left(\vec p_4 - C (\vec \ell_1-\vec \ell_2)\times(\vec \ell_3-\vec \ell_2)\right)\;
e^{iC \vec \ell_1 \cdot (\vec \ell_2 \times \vec \ell_3)},
\eea
where we have introduced a normalization factor ${\cal N}$.
Carrying out the integration over $\vec \ell_i$'s, we find
\be
{\cal A}_0 = \delta^{(3)}(\vec p_1 + \vec p_2 + \vec p_3 + \vec p_4) \; {\cal M}_0, 
\ee
where
\be \label{A}
{\cal M}_0 = 
2 \cos(C^{-1/2} \sqrt{|\vec p_1 \cdot (\vec p_2 \times \vec p_3)|}) \;
\frac{{\cal N}(\vec \ell)}{C^{9/2}|\vec p_1 \cdot(\vec p_2\times \vec p_3)|^{3/2}}.
\ee
It follows that we should choose
\be \label{N}
{\cal N}(\vec \ell) \equiv C^9 |\vec \ell_1 \cdot (\vec \ell_2 \times \vec \ell_3)|^3 
= C^{9/2}|\vec p_1 \cdot(\vec p_2\times \vec p_3)|^{3/2} \equiv {\cal N}(\vec p),
\ee
so that (\ref{A}) is just the sum over the interaction vertex (\ref{4v})
for the two possible orientations of the tetrahedron.

\begin{figure}[ptb]
\setlength{\unitlength}{3pt}
\par
\begin{center}
$\begin{array}{ccc}
\begin{picture}(40,30)(-20,-15)
\put(0,0){\circle{40}}
\put(-15,-6.8){\line(1,0){30}}
\put(-13,-10){$\vec p$}
\put(13, -10){$\vec p$}
\put(0,8){$\vec q$}
\end{picture} 
&
\begin{picture}(40,30)(-20,-15)
\put(0,0){\circle{40}}
\put(-16.8,10){\vector(1,-1){10}}
\put(-16.8,-10){\vector(1,1){10}}
\put(6.8,0){\vector(1,1){10}}
\put(6.8,0){\vector(1,-1){10}}
\put(-16,5){$\vec p_1$}
\put(-16,-5){$\vec p_2$}
\put(14,5){$\vec p_3$}
\put(14,-5){$\vec p_4$}
\end{picture}
&
\begin{picture}(40,30)(-20,-15)
\put(-15,0){\vector(1,0){30}}
\put(0,0){\circle{40}}
\put(-12,2){$\vec p$}
\put(10,2){$\vec p$}
\end{picture} \\
\mbox{(a)} & \mbox{(b)} & \mbox{(c)}
\end{array}$
\caption{(a) The 1-loop diagram corresponding to eq. (\ref{A1}). 
(b) The 1-loop diagram corresponding to eq. (\ref{A1p}) 
(c) The 2-loop diagram corresponding to eq. (\ref{A2}).} 
\label{OneLoopTwoLoop}
\end{center}
\end{figure}

%%%%%%%%%%%%%%%%%%%%%%%%%%%%%%%%%%%%%%%%%%%%
\subsubsection{One-loop diagram} 

The 1-loop diagram in Fig. (\ref{OneLoopTwoLoop}) (a) 
is not modified by the phase factor depending on $C$ 
because two of the momenta of the 4-membrane interaction vertex 
is the same vector, 
and the volume of the tetrahedron must be zero. 
This diagram is thus exactly the same as in usual field theory
\be \label{A1}
{\cal M}_1 = \int d^3 \vec p \frac{1}{\vec p^2 + m^2}.
\ee

Another 1-loop diagram, the one in Fig (\ref{OneLoopTwoLoop}) (b), is less trivial.
It is 
\bea
{\cal A}'_1 &=& \int \prod_{i=1}^6 d^3 \vec \ell_i \; d^3\vec p \; d^3\vec q \;
\delta^{(3)}(\vec p_1 - C \vec \ell_2\times \vec \ell_1)\;
\delta^{(3)}(\vec p_2 - C \vec \ell_3\times \vec \ell_2)\;
\delta^{(3)}(\vec p - C \vec \ell_5\times \vec \ell_4)\;
\delta^{(3)}(\vec q - C \vec \ell_6\times \vec \ell_5)\cdot \nn \\
&& \delta^{(3)}(\vec p - C \vec \ell_3\times \vec \ell_1)\;
\delta^{(3)}(\vec q - C \vec \ell_1\times \vec \ell_2)\;
\delta^{(3)}(\vec p_3 - C \vec \ell_6\times \vec \ell_4)\;
\delta^{(3)}(\vec p_4 - C (\vec \ell_4 - \vec \ell_5)\times (\vec \ell_6 - \vec \ell_5))\cdot \nn \\
&& \frac{1}{\vec p^2 + m^2} \frac{1}{\vec q^2 + m^2} \;
{\cal N}(\vec p_1, \vec p_2, \vec p)\;
{\cal N}(\vec p_3, \vec p_4, \vec p)\;
e^{iC \vec \ell_1 \cdot(\vec \ell_2 \times \vec \ell_3)} \;
e^{iC \vec \ell_4 \cdot(\vec \ell_5 \times \vec \ell_6)},
\eea
where ${\cal N}(\vec p_1, \vec p_2, \vec p)$ is defined in (\ref{N}).
It is straightforward to find, 
up to the delta function imposing momentum conservation, 
the amplitude ${\cal A}'_1$ is, up to the delta function imposing energy-momentum conservation,
\be
\label{A1p}
{\cal M}'_1 = \int d^3 \vec p \; \frac{1}{\vec p^2 + m^2} \frac{1}{(\vec p_1+\vec p_2-\vec p)^2 + m^2}
2\cos(C^{-1/2}\sqrt{|\vec p \cdot(\vec p_1 \times \vec p_2)|})
2\cos(C^{-1/2}\sqrt{|\vec p \cdot(\vec p_3 \times \vec p_4)|}).
\ee
%%% 0108 %%%
Generically (if $\vec p_1 \times \vec p_2$ and $\vec p_3 \times \vec p_4$ are not parallel),
$\vec p$ has an effective cutoff in two directions due to the two $\cos$ factors. 
The integral is obviously finite.

If the spacetime dimension is 6, we replace the momenta by 6-vectors.
We would then have another two $\cos$ factors due to the nonassociativity in the other 3 dimensions.
The 6 dimensional integral of $\vec p$ is unbounded only in 2 directions, 
and thus the integral is finite.
The diagram has a logrithmic divergence in 12 dimensions.
%% 0108 %%

%%%%%%%%%%%%%%%%%%%%%%%%%%%%%%%%%%%%%%%%%%%%
\subsubsection{2-loop diagram} 

The 2-loop diagram in Fig. \ref{OneLoopTwoLoop} (c) 
can be obtained from Fig. \ref{OneLoopTwoLoop} (b) 
by identifying $\vec p_2$ with $\vec p_3$ and integrating over $\vec p_2$. 
Therefore
\be
\label{A2}
{\cal M}_2 = \int d^3 \vec p' \; d^3 \vec q \; 
\frac{1}{\vec p'{}^2 + m^2} \frac{1}{\vec q^2 + m^2} \frac{1}{(\vec p-\vec p'-\vec q)^2 + m^2}
4\cos^2(C^{-1/2}\sqrt{|\vec p \cdot(\vec p' \times \vec q)|}).
\ee

The cosine factor imposes an effective cutoff at the energy scale 
\be
\Lambda_{\mbox{eff}} = \sqrt{\frac{C}{|p|}}\,.
\ee
As a result the integral is finite, 
although it would have a logrithmic divergence without the cosine factor.

The diagram diverges if the spacetime dimension is 6 or higher.

We suspect that there is no real divergence which can not be removed by normal ordering
in this scalar field theory living on a 3 dimensional nonassociative space.
(The divergence of ${\cal M}_1$ in (\ref{A1}) can be removed by normal ordering.)
%% p %%
% I think the following statement is wrong.
%In contrast, the 4-point interaction is non-renormalizable for field theory on ordinary space.

%%%%%%%%%%%%%%%%%%%%%%%%%%%%%%%%%%%%%%%%%%%%
\section{Inclusion of internal degree of freedom}

%%% 0108 %%%
In the above we have considered open membranes ending on a single M5-brane. 
For a stack of $N$ M5-branes, 
open membranes should acquire internal degrees of freedom analogous to Chan-Paton factors. 
Naively, there are cylindrical membranes stretched between 2 M5-branes, 
just like open strings stretched between 2 D-branes, 
and so one expects a non-Abelian gauge theory 
on the M5-brane world-volume at low energy.
However, anomaly and entropy computations \cite{N3} suggest that 
the world-volume theory has $N^3$ degrees of freedom,
rather than $N^2$ as in Yang-Mills theory. 
Interestingly, 
$N^3$ is precisely what our truncated model of open membrane would suggest.
%% PMH 0306 %%
\footnote{
The $N^3$ degrees of freedom can also be accounted for \cite{Berman:2006eu}
using the notion of fuzzy $S^3$ \cite{Ramgoolam:2001zx}, \cite{Ramgoolam:2002wb}.
}
%% PMH 0306 %%
For the triangular membrane, it is natural to introduce Chan-Paton factors 
either on the corners or on the edges. 
The Chan-Paton factors on the corners are easily incorporated in 
our formulation above, 
since it is simply the extension of the coordinates $\vec x_i$ ($i = 1,2,3$)
to include additional indices $(\vec x_i, n_i)$ $(i = 1,2,3)$.
Hence we will only discuss other less trivial ways of introducing internal degrees of freedom.

First we present a prescription \cite{Ambjorn_book, DePietri:2000ii}
which was originally proposed as a generalization
of the matrix model whose Feynman rule
produces the 3d gravity and fit with our formulation.
It corresponds to open membranes with 
boundaries divided into 3 sections belonging to 3 M5-branes. 
%% 0108 %%

We introduce three indices to each wave function
assigned with a triangle,
\be
\Phi(\vl)\quad \rightarrow\quad \Phi_{ijk}(\vl)
\ee
where the indices $i,j,k$ take values from $1,\cdots,N$.
We may assign the symmetry under the permutation of
indices.  For example, a natural generalization of the
Hermitian matrix is
\be
\mathrm{Re} (\Phi_{i_{\sigma(1)}i_{\sigma(2)}i_{\sigma(3)}})
=\mathrm{Re} (\Phi_{i_1 i_2 i_3})\,,\quad
\mathrm{Im} (\Phi_{i_{\sigma(1)}i_{\sigma(2)}i_{\sigma(3)}})
=(-1)^\sigma \mathrm{Im} (\Phi_{i_1 i_2 i_3})\,,\quad
\sigma\in S_3\,.
\ee
In either case, we assume that $\Phi_{ijk}$ is invariant
under the cyclic permutation of $i,j,k$.

We may attach the three indices to the three edges
of the triangle.  At the vertex, as our $\vl$,
we put the four field $\Phi^{(a)}$ at the faces of a
tetrahedron and identify the indices which share the 
same edge and sum over them.  More explicitly we have,
\be\label{quar2}
\left<\!\!\left<\Phi^{(1)},\Phi^{(2)},\Phi^{(3)},\Phi^{(4)}\right>\!\!\right>
=\sum_{i_1\cdots i_6=1}^N
\left<\Phi^{(1)}_{i_1,i_2,i_6},\Phi^{(2)}_{i_2,i_3,i_4},
\Phi^{(3)}_{i_3,i_1,i_5},\Phi^{(4)}_{i_4,i_5,i_6}\right>,
%\left<\Phi^{(1)}_{i_3,i_1,i_2},\Phi^{(2)}_{i_2,i_5,i_4},
%\Phi^{(3)}_{i_5,i_1,i_6},\Phi^{(4)}_{i_4,i_6,i_3}\right>,
\ee
where the single bracket $\left<\Phi,\Phi,\Phi,\Phi\right>$
is the quartic product (\ref{quartic}) which was defined by the integral over $\vl$.
The connections of colors at the vertex is the same as
figure \ref{fig:tetrahedra}.  
%Twists at $\Phi^{(2)},\Phi^{(4)}$ are needed to 
%keep the orientations of the triangle attached at the tetrahedron.
%%% pm %%%
% The labels in the figure does not seem to match the equation above
%
Similarly, the triplet product should be defined as
\be
\left[\!\!\left[\Phi^{(1)},\Phi^{(2)},\Phi^{(3)}\right]\!\!\right]_{i_4, i_5, i_6}
=\sum_{i_1\cdots i_3=1}^N
\left[\Phi^{(1)}_{i_1,i_2,i_6},\Phi^{(2)}_{i_2,i_3,i_4},
\Phi^{(3)}_{i_3,i_1,i_5}\right].
\ee

%%%%%%%%%%%%%%%%%%%%%%%%%%%%%%%%%%%%%%%%%%%%%%%%%%%%%
% \begin{figure}[bpt]
% \begin{center}
% \includegraphics[scale=0.5]{vertex.eps}
% \end{center}
% \caption{Connections of color indices
% at the vertex (the letter $i$ is omitted)}
% \label{fig:vertex}
% \end{figure}
%%%%%%%%%%%%%%%%%%%%%%%%%%%%%%%%%%%%%%%%%%%%%%%%%%%%%

\begin{figure}[bpt]
\begin{center}
\includegraphics[scale=0.9]{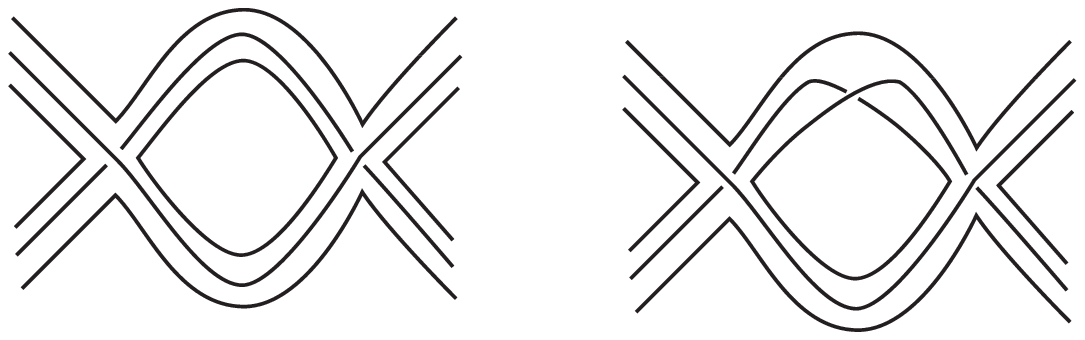}
\end{center}
\caption{Two fat graphs associated with Feynman graph 
Fig.\ref{OneLoopTwoLoop}(b)}
\label{fig:fat_graphs}
\end{figure}

By writing three lines along each propagator and 
the connections of these lines at the vertices,
one can write a {\em fat graph} associated with 
a given Feynman diagram. As an example, in fig. \ref{fig:fat_graphs},
we give two types
of the fat graphs associated with the Feynman graph fig. \ref{OneLoopTwoLoop}b.
As the diagrams of usual gauge theories,
one obtains the factor of $N$ for each loop of the index line.
Since one may use twist $\sigma\in S_3$ for
the propagator, there are a few different fat graphs associated
with a single Feynman diagram and each graph may have
different dependence on $N$. In the fig. \ref{fig:fat_graphs},
the left graph has factor $N$ from the index loop.  On the other
hand, the right graph does not have that factor and is $O(1)$.
This is very similar to the
computation of the gauge theory.  For the gauge theory,
if we forget the dependence on $x$, the system is reduced
to the matrix model.  The dual diagram of the fat
graph corresponds to the triangulation of the
2d surface.  A similar phenomena happens in our case too.
If we forget the $\vl$ dependence, the Feynman integral
reduces to the integration over the cubic matrix $\Phi_{ijk}$.
We can associate each vertex an tetrahedron
and the propagator indicates how two tetrahedra are connected
at the faces.  In this way, the dual diagram to each fat graph
gives a symplicial decomposition of the 3d space.
In \cite{DePietri:2000ii}, it was shown that the two parameters
$g$ and $N$ are identified with the %coefficient of the Einstein-Hilbert action 
Newton's constant and the consmological constant. 
In the 3d case, however, there is the difficulty that
the dual graph for the Feynman graph may not define a
manifold.  So the correspondence between the Feynman graphs
and the lattice gravity is not so clear as the 2d case.

%%%%%%%%%%%%%%%%%%%%%%%%%%%%%%%%%%%%%%%%%%%%%%%%%%%%%
%%%%%%%%%%%%%%%%%%%%%%%%%%%%%%%%%%%%%%%%%%%%%%%%%%%%%

\paragraph{Other possibilities}
Here we examine a few more ideas to introduce the
internal degree of freedom.
One such possibility is to consider an analog of the noncommutative
torus with the rational $\theta$ parameter.
It is equivalent to the matrix algebra. We recall that the
basis of $N\times N$ matrix is spanned by the matrices
$U^{n_1} V^{n_2}$ ($n_1, n_2\in \mathbf{Z}_N$) where
$U,V$ satisfies $UV = VU e^{2\pi i/N}$. It is then easy to see that
\be
(U^{n_1}V^{n_2})( U^{m_1} V^{m_2})
= e^{-\frac{2\pi i}{N}(n_1m_2-m_1 n_2)}(U^{m_1} V^{m_2})(U^{n_1}V^{n_2})\,.
\ee
Therefore the matrix algebra is expressed as the noncommutative phase
$(n_1,m_1; n_2, m_2)\rightarrow e^{\frac{2\pi i}{N}(n_1m_2-m_1 n_2)}$.

One may generalize this idea to three dimensions.
We extend $\mathbf{R}^3$ to the product space
$\mathbf{R}^3 \times \mathbf{Z}_N^{\otimes 3}$, i.e.,
we require that the field $\Phi$ depends on these extra ``dimensions''
$\Phi(\vl_1,\vl_2,\vl_3;\vn_1, \vn_2, \vn_3)$
($\vn_i\in \mathbf{Z}_N^{\otimes 3}$, $\vn_1+\vn_2+\vn_3\equiv0$ mod $N$).
The quartic product is then defined as (after neglecting degree of
freedom of $\vl$)
\bea
\left<\Phi^{(1)},\Phi^{(2)},\Phi^{(3)},\Phi^{(4)}\right>
&=& \sum_{\vn_1\cdots\vn_6\in \mathbf{Z}_N} 
\,e^{\frac{2\pi i}{N}\vn_1\cdot(\vn_2\times\vn_3)}\,\cdot
\nonumber\\
&&\cdot\hat\Phi^{(1)}(\vn_1,\vn_6, -\vn_2)
\hat\Phi^{(2)}(\vn_2,\vn_4, -\vn_3)
\hat\Phi^{(3)}(-\vn_1,\vn_3, \vn_5)
\hat\Phi^{(4)}(-\vn_4,-\vn_6, -\vn_5)\cdot\nonumber\\
&&\cdot \,
\delta^{(3)}(\vn_1+\vn_6-\vn_2)
\delta^{(3)}(\vn_2+\vn_4 -\vn_3)
\delta^{(3)}(-\vn_1+\vn_3+ \vn_5)\,.
\eea

As in the relation between $\vec p$ and $\vl$,
the ``ground state'' wave function may depends only on 
$\lambda=\vn_1\times \vn_2$,
\be
\Phi(\vn_1,\vn_2,\vn_3)=\psi(\vn_1\times\vn_2)\,.
\ee
In this case, the quartic product 
$[\Phi^{(1)},\Phi^{(2)},\Phi^{(3)},\Phi^{(4)}]^{(4)}$
is simplified to,
\bea
\sum_{\vlam_1,\cdots,\vlam_4\in \mathbf{Z}_N}
W(\vlam)\delta(\sum_{i=1}^4\vlam_i)
\cos\left({\frac{2\pi }{N}\sqrt{|\vlam_1\cdot(\vlam_2\times \vlam_3)|}}\right)
 \psi_1(\vlam_1)
\psi_2(\vlam_2)
\psi_3(\vlam_3)
\psi_4(\vlam_4),
\eea
where 
\bea
W(\vlam)&=&\sum_{\vn_1\cdots\vn_6\in \mathbf{Z}_N}
\delta(\vlam_1-\vn_1\times\vn_2)
\delta(\vlam_2-\vn_2\times\vn_3)
\delta(\vlam_3-\vn_3\times\vn_1)
\delta(\sum_{i=1}^3\vlam_i-\vn_4\times\vn_5)\cdot\nonumber\\
&&\cdot\delta^{(3)}(\vn_1+\vn_6-\vn_2)
\delta^{(3)}(\vn_2+\vn_4 -\vn_3)
\delta^{(3)}(-\vn_1+\vn_3+ \vn_5)\,.
\eea
is the non-negative integer weight for $\vlam$.
When the square root phase factor 
$\sqrt{|\vlam_1\cdot(\vlam_2\times \vlam_3)|}$
becomes non-integer for generic $\lambda$, this weight
factor vanishes.
Unlike the two dimensional case, this assignment
of the internal degree of freedom is not equivalent to
the previous one.

%%%%%%%%%%%%%%%%%%%%%%%%%%%%%%%%%%%%%%%%%%%%%
\section{Some generalization of
membrane fields and their interactions} \label{general} 

In this paper, we concentrate ourselves to the membrane
field which is represented by a triangle.  This restriction
is chosen in order to obtain the simplest toy model
to describe the basic feature of the membrane theory.

It is, however, obvious that the membranes can take
arbitrary shape even if we restrict them to the plane perpendicular
to the momentum.  It implies that we miss 
most of the interactions except for the simplest one.
If we allow them to take arbitrary shape, however, we would
not be able to escape from the instability.  So as the next 
approximation, one may have milder restriction to
 the shape of the membrane 
to the polygons. The membrane field associated with
$n$-polygon $\Delta$ is written as a function of its
$n$ vertices as
$
\Phi(\vec x(\Delta))
$
where we introduced a shorthand notation,
$x(\Delta)=(\vec x_1,\cdots, \vec x_n)$.
We assume that $n$ vectors
$\vec x_1,\cdots,\vec x_n$ are located on the same plane.
As before, we need to require that the momentum is proportional
to the area vector,
\be
\vec p -C\vec A(\Delta)=0\,,\quad
\vec A(\Delta)=\frac{1}{2}\sum_{1\leq i<j\leq n} \vl_i\times \vl_j\,,
\ee
where
$
\vl_i=\vec x_{i+1}-\vec x_i$, ($\vec x_{n+1}\equiv \vec x_1$).
In terms of the membrane field, we can write it as
\be
\tilde \Phi(\vec p;\vl(\Delta))=\hat \Phi(\vl(\Delta))
\delta^{(3)}(\vec p-C \vec A(\Delta))
\ee
where $\tilde \Phi(\vec p;\vl(\Delta))$ is the Fourier
transformation of $\Phi(\vec x(\Delta))$,
\be
\tilde \Phi(\vec p;\vl(\Delta))
= \int d\vec x_c^\Delta e^{i\vec p \cdot \vec x_c^\Delta}
\hat\Phi(\vec x(\Delta))\,,
\quad
\vec x_c^\Delta=\frac{1}{n}\sum_{i=1}^n \vec x_i\,,\quad
\vl(\Delta)=(\vl_1,\cdots,\vl_n)\,. 
\ee
Here we need to change the variables from $\vec x_i$ to
$\vec p$ and $\vl_i$. 
The ground state of the membrane field is described by
the wave function whose $\hat\Phi$ depends only on 
the momentum (or the area vector),
\be
\hat\Phi(\vl(\Delta))=\phi(\vec p)=\phi(C \vec A(\Delta))\,.
\ee

The interaction among the generalized membrane fields
can be defined as before.  We introduce a three dimensional
open space $B$ (which may not necessarily have the topology of
ball) whose boundary $\partial B$ is covered by
polygons $\Delta_f$ ($f=1,\cdots, b_2$), 
$\partial B=\cup_f \Delta_f$.  We write
the number of edges and vertices of $\partial B$
as $b_1$ and $b_0$ respectively. The membrane vertex
associated with $B$ is then written as
\be
\left<\Phi_1,\cdots, \Phi_{b_2}\right>_B=
\int d \vec x\,
\Phi_1(\vec x(\Delta_1))\cdots \Phi_{b_2}(\vec x(\Delta_{b_2}))
\ee
where the integration is over the all vertices on $\partial B$.
$\vec x$'s which belong to the same $\Delta$ should be
located on the same plane.

In terms of the $\hat \Phi$ components, this product
is written as
\be
\delta\left(\sum_{i=1}^{b_2}\vec p_i\right)
\int d \vl\,
e^{i C V(B)}
\hat \Phi(\vl(\Delta_1))\cdots
\hat\Phi(\vl(\Delta_{b_2})),
\ee
where $V(B)$ is the volume of $B$.

We note that we can add arbitrary type of the interaction
vertices.  So far, we do not have a general principle for
how to select them.  As in string field theory, we need some sort of 
gauge princple.  This is outside of the scope of the current paper
but will be the most important issue in future study.

We note that as $n\rightarrow \infty$, our membrane field
is getting closer to the closed string field
$\Phi[X(\sigma)]$ living on the plane determined
by the momentum $\vec p$. In this sense,
our membrane field theory has some
connection with (discretized) closed string field theory
\cite{Zwiebach:1992ie}.

We can include  the color degree of freedom to the
membrane field associated with the polygon similary.
The simplest way is to introduce the color index
$i$ ($i=1,\cdots, N$) to each edge of the polygon.
For $n$-gons, we have $n$ color indices.
We can assume the symmetry with respect to
the cyclic permutation of these indices as before.
At the vertex, since each edge is always shared
by two faces, we can define the identification 
between two indices associated with it.

\paragraph{Reduction to the noncommutative field theory}
As an application of the use of general membrane field,
we would like to derive the noncommutative field
theory from the membrane through double
dimensional reduction.
%%% 0108 %%% although it is rather trivial.%% 0108 %%

We assume that a spacial direction (say $x^3$) is compactified as
$x^3 \sim x^3+2\pi R$ and every membrane wraps this direction
once.  We suppose the $C$ field takes the form 
$C_{ijk}=C \epsilon_{ijk}$ ($i,j,k=1,2,3$) as usual.
Since the membrane should wrap the $3$-direction,
the simplest shape it can take is the cylinder,
which can be spacified by two vectors,
$\vl_1$ in $1-2$ directions and
$\vl_2=2\pi R \vec e_3$ ($\vec e_3$ is the unit
vector in the $3$-direction.
With this set-up, the momentum vector
$\vec p$ should be in $1-2$ directions and
the relation $\vec p= C\vl_1\times \vl_2$
becomes the standard relation for the
noncommutative geometry,
%% YM 0305 %%
\be
p_i= B \epsilon_{ij} (\ell_1)_j\,,\quad
i,j=1,2\,,\quad
B=2\pi R C\,.
\ee
%% YM 0305 %%
The simplest interaction among such membrane fields
are given by the three-point function.
The topology of the interaction vertex is given
by a solid torus whose boundary is the torus
which is divided into three cylinders attached to
the membrane fields.  The volume of the
solid torus is given by $\pi R C |p_1\times p_2|
= \frac{1}{2} B |p_1\times p_2|$.  This is the
phase factor of the noncommutative field theory.

Since the cylinder has two edges, the number of 
color degree of freedom is two.  This is consistent with
the noncommutative field theory whose internal degree of
freedom is given by the matrix.

\section{Connection with lattice gravity}
\label{sec:higher}
In this section we make some comments on the
relation between our model with lattice 
quantum gravity. It appears through the 
consistency conditions for the products
among the membrane fields.

In 2d (the noncommutative case), higher vertices
can be constructed by using the star product
as $\mbox{Tr}(\Phi_1\star\cdots \star \Phi_n)$,
which defines the polygon with $n$ edges.
The consistency condition of the construction of such
vertices is that the quantity defined above does not
depend on the order of taking the star product.  This is
exactly the associativity condition of the star product.

For the 3d case, we have a different type of product
(the triplet product) which is defined for
three fields instead of two.  So we need to find the consistency
condition which replaces the associativity.
We would like to argue that this may be identical with the
consistency conditions of 3d lattice gravity.
The connection with the lattice gravity seems  natural
since the membrane theory should contain the 3d gravity
as the dynamics of the world-volume.

%%%%%%%%%%%%%%%%%%%%%%%%%%%%%%%%%%%%%%%%%%%%%%%%%%%%%
\subsection{2d (noncommutative) case}
In the noncommutative case, the field $\Phi$ depends only on one momentum
$\vec k$.  We consider the consistency condition when
we compute four-point functions
\be
\mbox{Tr} \left(\Phi_1\star
\Phi_2
\star\Phi_3
\star\Phi_4\right)
=\int d^4\vec k
\Phi_1(\vec k_1)\star
\Phi_2(\vec k_2)
\star\Phi_3(\vec k_3)
\star\Phi_4(\vec k_4)
\delta(\vec k_1+\vec k_2 +\vec k_3+\vec k_4)\,.
\ee
There are two different ways to evaluate this quantity.
(1) We take the star product $\Phi_1\star \Phi_2$ and
$\Phi_3\star \Phi_4$ and take their inner product.
(2) Similarly we take the star products $\Phi_2\star \Phi_3$ and
$\Phi_4\star \Phi_1$ and take their inner product.
These two computations match because of the associativity
of the star product.   More explicitly,
\bea
\mbox{(1)}&=&
\int d^4\vec k d\vec k_5
\Phi_1(\vec k_1)
\Phi_2(\vec k_2)e^{i\theta \vec k_1\times \vec k_2}
\delta(\vec k_1+\vec k_2+\vec k_5)\cdot\nonumber\\
&&~~~~~~\cdot 
\Phi_3(\vec k_3)
\Phi_4(\vec k_4)e^{i\theta \vec k_3\times \vec k_4}
\delta(\vec k_3+\vec k_4-\vec k_5)
\delta(\vec k_1+\vec k_2 +\vec k_3+\vec k_4)\nonumber\\
&=&\int  d^4\vec k d\vec k_6
\Phi_2(\vec k_2)
\Phi_3(\vec k_3)e^{i\theta \vec k_2\times \vec k_3}
\delta(\vec k_2+\vec k_3+\vec k_6)\cdot\nonumber\\
&&~~~~~~\cdot
\Phi_4(\vec k_4)
\Phi_1(\vec k_1)e^{i\theta \vec k_4\times \vec k_1}
\delta(\vec k_4+\vec k_1-\vec k_6)
\delta(\vec k_1+\vec k_2 +\vec k_3+\vec k_4)\nonumber\\
&=&\mbox{(2)}\,.
\eea
The equality comes from the fact that the phase satisfies
$
e^{i\theta \vec k_1\times \vec k_2+i\theta \vec k_3\times \vec k_4}
=e^{i\theta \vec k_2\times \vec k_3+i\theta \vec k_4\times \vec k_1}
$
when $\sum_{i=1}^4 \vec k_i=0$. 
We refer to this condition as the consistency condition for 
the $2\leftrightarrow 2$ move (fig.\ref{NonCommConsistency}a).

\begin{figure}[ptb]
\setlength{\unitlength}{3pt}
\par
\begin{center}
$\begin{array}{cc}
\begin{picture}(40,30)(-5,-20)
% draw tetrahedron
\put(15,10){\vector(-3,-2){15}}
\put(3,5){$\vec k_1$}
\put(0,0){\vector(3,-2){15}}
\put(5,-7){$\vec k_2$}
\put(30,0){\vector(-3,2){15}}
\put(23,5){$\vec k_4$}
\put(15,-10){\vector(3,2){15}}
\put(25, -7){$\vec k_3$}
\put(15,-10){\line(0,1){20}}
\multiput(0,0)(5,0){6}{\line(1,0){3}}

\end{picture} 
&
\begin{picture}(40,30)(-10,-10)
\put(16,24){\vector(-2,-3){16}}
\put(28,0){\vector(-1,2){12}}
\put(22,12){$\vec k_2$}
\put(4,12){$\vec k_1$}
\put(0,0){\vector(1,0){28}}
\put(14,-5){$\vec k_3$}
{\color{Dark}
\put(16,24){\vector(0,-1){16}}
\put(12,12){$\vec k_5$}
\put(0,0){\vector(2,1){16}}
\put(7,5){$\vec k_4$}
\put(28,0){\vector(-3,2){12}}
\put(20,5){$\vec k_6$}
}
\end{picture} \\
\mbox{(a)} & \mbox{(b)}
\end{array}$
\caption{(a) $2\leftrightarrow 2$ move. 
(b) $1\leftrightarrow 3$ move.} 
\label{NonCommConsistency}
\end{center}
\end{figure}

In this case, this condition is sufficient to make the perturbation theory
consistent.  In the context of two dimensional gravity, however,
it is more natural to include the  invariance by dividing one triangle
into three. To describe it, we introduce a notation,
\bea
&&[\Phi]^{(2)}(\vec k_1, \vec k_2, \vec k_3)=\Phi(\vec k_1)
e^{i\theta \vec k_1\times \vec k_2}\qquad 
(\vec k_3=-\vec k_1-\vec k_2)\,,\\
&&([\Phi_1]^{(2)},[\Phi_2]^{(2)},[\Phi_3]^{(2)})
=\int d\vec k_1\cdots d\vec k_6
[\Phi_1]^{(2)}(\vec k_1,\vec k_4, -\vec k_5)
[\Phi_2]^{(2)}(\vec k_2,\vec k_5, -\vec k_6)
[\Phi_3]^{(2)}(\vec k_1,\vec k_6, -\vec k_4)\cdot
\nonumber\\
&&~~~~~~~~~~~~~~
\cdot
\delta(\vec k_1+\vec k_2+\vec k_3)
\delta(\vec k_1+\vec k_4-\vec k_5)
\delta(\vec k_2+\vec k_5-\vec k_6)
\delta(\vec k_3+\vec k_1-\vec k_4).
\eea
The consistency for the $1\leftrightarrow 3$ move is stated as
\be
\mbox{Tr}(\Phi_1\star \Phi_2\star \Phi_3)
=([\Phi_1]^{(2)},[\Phi_2]^{(2)},[\Phi_3]^{(2)})\,.
\ee

%%%%%%%%%%%%%%%%%%%%%%%%%%%%%%%%%%%%%%%%%%%%
\subsection{3d (nonassociative) case} \label{gravity}
In order to generalize the idea of the 2d case to 3d, it
is natural to consider the vertices which are defined
for arbitrary  triangulated 3d (open) bodies $B$. Namely,
they are decomposed as
\be
B=\cup_a \Delta_a\,,\quad
\partial B= \cup_q \lambda_q\,,\quad
\lambda \in \partial \Delta_a \quad
\mbox{for a $\Delta_a$}
\ee
where $\Delta_a$ are tetrahedra and $\lambda_q$
are triangles. In order to define the amplitude
for $B$, we assign a factor $W(\Delta_a)$ for each tetrahedron
and multiply them.  Schematically they are written as
\be
\left<\Phi_1,\cdots,\Phi_{b_2}\right>=\sum_{\vl_e,\,e\in\mathrm{edges}}{}'\left(
\prod_{t\in \mathrm{tetrahedra}} W(\Delta_t)
\prod_{f\in \mathrm{faces\ on\ }\partial B} \Phi_f(\Delta_f)
\right)\,.
\ee
where $\sum$ is the summation over the edges
around the internal triangles where
two $\Delta_a$ are overlapped.  
By the prime in the summation, we imply that the variables
assigned to the edges ($\vl$'s) are constrained as the difference
between the locations of the corners ($\vec x$'s) which are connected
by the edge.  This is different from the situation in 3d gravity
where the summation of the edge variables are more free.
% The triangles at the
% boundary should be then integrated after multiplying the
% wave function which should be attached to each $\lambda_q$.

In 3d gravity, we should impose the condition that
$\left<\Phi_1,\cdots,\Phi_{b_2}\right>$ 
would not depend on the triangulation.
This consistency condition is summarized by two conditions:
invariance under the $2\leftrightarrow 3$ move and $1\leftrightarrow 4$ 
move (fig.7).

\begin{figure}[tbp]
\setlength{\unitlength}{3pt}
\par
\begin{center}
$\begin{array}{cc}
\begin{picture}(40,45)(0,-25)
% draw tetrahedron
\put(0,-0.2){\line(4,-1){40}}
\put(40,-10){\line(0,1){20}}
\put(39.8,-10){\line(0,1){20}}
\put(20,20){\vector(-1,-1){20}}
\put(20,19.8){\vector(-1,-1){20}}
\put(10, 10){$\vec \ell_1$}
\put(20,-20){\vector(-1,1){20}}
\put(20,-20.2){\vector(-1,1){20}}
\put(10, -10){$\vec \ell_4$}
\put(20,20){\vector(2,-3){20}}
\put(20,19.8){\vector(2,-3){20}}
\put(25,10){$\vec \ell_2$}
\put(20,20){\vector(2,-1){20}}
\put(20,19.8){\vector(2,-1){20}}
\put(30,15){$\vec \ell_3$}
\put(20,-20){\vector(2,1){20}}
\put(20,-20.2){\vector(2,1){20}}
\put(30,-16){$\vec \ell_5$}
{\color{Dark}
\put(0,0){\line(4,1){40}}
\put(0,0){\line(4,-1){40}}
\put(20,-20){\vector(2,3){20}}
\put(24,-14){$\vec \ell_6$}
\multiput(20,-20)(0,5){8}{\line(0,1){4}}
}

\end{picture} 
&
\begin{picture}(40,45)(-10,-15)
% draw tetrahedron
\put(16,24){\vector(-2,-3){16}}
\put(16,24.15){\vector(-2,-3){16}}
\put(4,12){$\vec \ell_1$}
\put(16,24){\vector(1,-2){12}}
\put(16,24.2){\vector(1,-2){12}}
\put(19, 5){$\vec \ell_2$}
\put(16,24){\vector(1,-4){8}}
\put(16,24.2){\vector(1,-4){8}}
\put(24, 10){$\vec \ell_3$}
\put(0,0){\line(3,-1){24}}
\put(0,0.2){\line(3,-1){24}}
\put(28,0){\line(-1,-2){4}}
\put(28,0.2){\line(-1,-2){4}}
{\color{Dark}
\put(0,0){\line(1,0){28}}
\put(16,24){\vector(0,-1){20}}
\put(15,8){$\vec \ell_5$}
\put(16,4){\line(-4,-1){16}}
\put(16,4){\line(2,-3){8}}
\put(16,4){\line(3,-1){12}}
}
\end{picture} \\
\mbox{(a)} & \mbox{(b)}
\end{array}$
\caption{(a) $2\leftrightarrow 3$ move. 
(b) $1\leftrightarrow 4$ move.} 
\end{center}
\end{figure}

In our case, the factor associated with each tetrahedron
$W(\Delta_t)$ takes the form
\be
W(\Delta_t)
=\exp\left(iC
\vec\ell_1\cdot(\vec\ell_2\times\vec\ell_3)
\right)=
e^{iCV(\Delta_t)}
\ee
where $V(\Delta_t)$ is the volume of $\Delta_t$.
So the above consistency conditions are nothing but
the requirement that the volume does not
depend on the way to divide it into tetrahedra.
More explicitly,
\begin{eqnarray}
 &&  W(\vec\ell_1,\vec\ell_2,\vec\ell_3)
 W(-\vec\ell_4,-\vec\ell_5,-\vec\ell_6)\nonumber\\
&&~~~~= W(\vec\ell_1,\vec\ell_2,\vec\ell_1-\vec\ell_4)
W(\vec\ell_2,\vec\ell_3,\vec\ell_2-\vec\ell_5)
W(\vec\ell_3,\vec\ell_1,\vec\ell_3-\vec\ell_6)
\label{2-3}
\nonumber\\
&&~~~~\mbox{if}\quad
\vec\ell_2-\vec\ell_1=\vec\ell_5-\vec\ell_4,\quad
\vec\ell_2-\vec\ell_3=\vec\ell_6-\vec\ell_5 , \\
&& W(\vec\ell_1,\vec\ell_2,\vec\ell_3)
=W(\vec\ell_1,\vec\ell_2,\vec\ell_5)
W(\vec\ell_2,\vec\ell_3,\vec\ell_5)\cdot
\nonumber\\
&&~~~~~~~\cdot W(\vec\ell_3,\vec\ell_1,\vec\ell_5)
W(\vec\ell_2-\vec\ell_3,\vec\ell_1-\vec\ell_3,\vec\ell_5-\vec\ell_3).
\label{1-4}
\end{eqnarray}
While these are rather trivial statements, it is a natural
generalization of the consistency conditions in two dimensions.

{}After this observation, we arrive at another interesting
possibility to introduce the internal degree of freedom
which satisfies the constraints above.  Namely, we multiply
the factor $W(\Delta)$ with the weight factor of 
lattice 3d gravity where $W(\Delta)$ is written in
terms of the  $q$-deformed $6j$ symbol of $SU(2)$.
\be
W(\Delta)\propto \left\{
\begin{array}{c c c}
j_1 & j_2 & j_3\\
j_4 & j_5, & j_6
\end{array}
\right\}_q
\ee
where $j$'s are assigned to the edges.
With this factor, one may introduce the effect of quantum
fluctuation of the world volume of the membrane.

%%%%%%%%%%%%%%%%%%%%%%%%%%%%%%%%%%%%%%%%%%%
\section{Outlook}

In this paper we studied open membranes in the large $C$-field background, 
%(we can also call them topological open membranes or closed strings),
mostly by making analogy with open strings in the large $B$-field background. 
We find that
\begin{itemize}
\item the 4-membrane interaction induces a triplet product which 
is a deformation of the Nambu bracket. 
%\item the algebra of the triplet product can be viewed as a nonassociative algebra.
\item the 4-membrane interaction naturally leads to the notion of truncated open membranes. 
\item the 4-membrane interaction (of ground states in $C$-field background) 
agrees with the natural definition of 4-membrane interaction of truncated open membranes in generic backgrounds. 
\item the triplet product satisfies consistency conditions for 3d lattice gravity, 
and a higher vertex composed of tetrahedra can be interpreted as a triangulation of the open membrane.
\item internal degrees of freedom analogous to Chan-Paton factors can be added to the boundaries of open membranes, 
and it naturally leads to the conclusion that there are $N^3$ degrees of freedom in the low energy effective field theory 
of M5-branes.
\item UV/IR mixing appears in a toy model of the effective field theory for open membrane ground states.
\end{itemize}

% Effective field theory of M5-branes

It will be interesting to study other aspects of nonassociative field theories, 
e.g., internal degrees of freedom, which we touched on briefly in this paper. 
One of the most important class of theories in physics is Yang-Mills theory, 
which is a gauge theory based on the noncommutative algebra. 
It will be very exciting if gauge symmetry can be generalized to 
the nonassociative algebra. 
This is a long-standing problem which has been associated with the Nambu bracket. 
A classical gauge symmetry without internal degrees of freedom 
is easy to construct \cite{Pioline:2002ba}. 
While area-preserving diffeomorphism can be realized via the Poisson bracket, 
one can realize volume-preserving diffeomorphism via the Nambu bracket (\ref{CNB})
\be \label{vpd}
\delta f(x) = \{ f(x), \a(x), \b(x) \},
\ee
where $(\a(x), \b(x))$ are infinitesimal gauge parameters. 
Let us define gauge potential $A_i(x)$ to transform as
\be
\delta A_i(x) = \{ x_i + A_i(x), \a(x), \b(x) \}.
\ee
Since the Nambu bracket satisfies the fundamental identity (\ref{FI1}), 
the field strength defined by 
\be
F(x) \equiv \{x_1 + A_1(x), x_2 + A_2(x), x_3 + A_3(x) \} - 1
\ee
transforms covariantly
\be
\delta F(x) = \{F(x), \a(x), \b(x)\}.
\ee
Assuming that the Leibniz rule is satisfied, 
the action 
\be
S = \int d^3 \vec x \frac{1}{2} F^2(x)
\ee
is gauge invariant. 

Let us recall that for the area-preserving diffeomorphism, 
one can quantize the Poisson bracket and 
approximate the area-preserving diffeomorphism by $U(N)$ (or $O(N)$ etc) symmetry.
It is then a happy coincidence that 
both the $U(1)$ gauge theory on a single noncommutative D-brane 
and the $U(N)$ gauge theory on $N$ D-branes 
have essentially the same algebraic structure of non-Abelian gauge theory. 

It is therefore desirable to have a discretized/deformed version of 
the volume-preserving diffeomorphism, 
so that it only involves $N^3$ (rather than infinitely many) degrees of freedom. 
Then we would expect that the low energy effective field theory of $N$ M5-branes 
would be a theory with this gauge symmetry. 

%%%  YM0111 %%%
We note that a small portion of the symmetry already exists 
in our set-up. In the definition of the inner product 
$(\!(\Phi^{(1)},\Phi^{(2)})\!)\equiv \sum_{ijk} (\Phi^{(1)}_{ijk})^*
\Phi^{(2)}_{ijk}$ and
the quartic product with colors (\ref{quar2}), 
it is easy to see that these products are invariant under 
global $O(N)$ rotation,
\be
\delta \Phi_{ijk}=\sum_{r} (\Phi_{rjk} \alpha_{ri}+
\Phi_{irk} \alpha_{rj}+\Phi_{ijr} \alpha_{rk})\,,\quad
\alpha_{ij}=-\alpha_{ji}\,.
\ee
This symmetry exists even for our definition of the product between
the generalized membrane fields associated with the
polygons since  we always attach polygons which are connected through
the edges (Sec. 7).  This symmetry is closely related to the
gauge symmetry of the string theory in the following sense.
As we have seen, the flat open membrane field on a polygon
is characterized by its boundary, namely the closed string.
On a circle, we may have $n$ different type of the boundary
condition if we insert twist fields at $n$ points.  
The closed string field with this set-up would
transform as mentioned above.  
%% YM 0305 %%
In the large $N$ limit, 
the $O(N)$ symmetry can be matched with the are-preserving diffeomorphism 
of the open membrane boundaries, 
in the same fashion that $U(N)$ symmetry is identified with 
the area-preserving diffeomorphism in the membrane theory
in the light cone gauge \cite{deWit:1988ig}.
%% YM 0305 %%

%Large $N$ limit

Furthermore, we would like to suggest the possibility of constructing a (cubic) matrix model 
for open membranes, 
analogous to the matrix model for noncritical strings. 
Ignoring spacetime coordinates and momenta and keeping only the internal degrees of freedom, 
our toy model becomes a model of cubic matrices.
Depending on how we introduce internal degrees of freedom, 
we obtain different types of 4-point interactions. 
A Feynman diagram with $n$ external legs can be understood as a triangulation of 
a 3d space with $n$ boundaries. 
It is also clear that the fat graphs of Feynman diagrams are sensitive to some 3d topological features.
It remains to be seen whether in a certain double scaling limit with $N\rightarrow \infty$,
the Feynman diagrams are dominated by those approximating a 3-sphere 
with $n$ boundaries. 

A related question of great importance is whether there is a limit analogous to 't Hooft's large $N$ limit
in which the importance of a Feynman diagram is characterized by a 3d topological invariant.

Another related subject is closed string field theory \cite{Zwiebach:1992ie}. 
Since we have so far completely ignored the interior of the open membrane, 
one can also interpret it as a closed string. 
The algebraic structure derived for truncated open membranes 
may be a good starting point to construct the full algebraic structure of closed string field theory, 
where we need $n$-products for all $n$ satisfying certain recursion relations. 
Now we only have triangular closed strings and the triplet product. 
After we extend the consideration to all polygons, 
we would naturally introduce more general products involving an arbitrary number of fields. 
It will be interesting to see whether such a lattice version of closed string 
also admits a simplified version of the algebraic structure of closed string field theory. 

%%%%%%%%%%%%%%%%%%%%%%%%%%%%%%%%%%%%%%%%%%%%%%%%%%%%
\noindent {\bf Acknowledgments}: 

We appreciate partial financial support from
Japan-Taiwan Joint Research Program 
provided by Interchange Association (Japan)
by which this collaboration is made possible. We also appreciate
the hospitality of the people in the National Taiwan University 
and the University of Tokyo where the most
of the work was done.

P.-M. H. thanks Chong-Sun Chu, Kazuyuki Furuuchi, Takeo Inami, Shunsuke Teraguchi, 
Wen-Yu Wen and Syoji Zeze.
He is supported in part by
the National Science Council,
and the National Center for Theoretical Sciences, Taiwan, R.O.C.
Y. M. is grateful to Y. Nambu, A. Hashimoto and
Sangmin Lee and S. Teraguchi for their interest
and the comments to this work.
He is partially supported by
Grant-in-Aid (\#16540232) from the Japan 
Ministry of Education, Culture, Sports,
Science and Technology.

\appendix

%%%%%%%%%%%%%%%%%%%%%%%%%%%%%%%%%%%%%%%%%%%%%%%%%%%%
\section{Some relations with nonassociativity
in open string, Nambu bracket and cubic matrix}

In this paper, we consider the open membrane theory
in the constant three form background $C$.
In the literature, this set-up is used to define
the nonassociative star product for the open string
degree of freedom and in many cases they are related
with Nambu bracket.  In this section, we provide a brief
summary of these studies in order to illuminate
the difference  or the correspondence of our approach with
them. Because of this nature, this section may be skipped
if the reader is not interested in the relation with previous works.  

\subsection{Nonassociative product in open string theory}

From open strings' viewpoint, the nonassociativity of
the star product appears when the $B$-field background
has non-trivial curvature, $C=dB \neq 0$.
In such situation, Kontsevich's star product \cite{Kontsevich:1997vb}
becomes nonassociative \cite{Cornalba:2001sm}
(while it is still possible to modify the product to be associative 
by including all phase space variables in a bigger algebra
\cite{Ho:2001qk}).
For the constant $C$-field background, 
the non-associativity of the algebra can be expressed in the form
\cite{Bouwknegt:2004ap}\cite{Ellwood:2006my},
\be \label{phiphiphi}
(\phi(\vec k_1)\star\phi(\vec k_2))\star \phi(\vec k_3)=
e^{iC \vec k_1\cdot (\vec k_2\times \vec k_3)}
\phi(\vec k_1)\star (\phi(\vec k_2)\star \phi(\vec k_3))
\ee
where 
\be
\phi(\vec k_1) \star \phi(\vec k_2)
=\pi(\vec k_1,\vec k_2) \phi(\vec k_1+\vec k_2)\,,\quad
\pi(\vec p,\vec q)=\exp\left(iC (2p_1+q_1+d)(p_2 q_3-p_3 q_2)\right)\,,
\ee
where $d$ is an extra parameter which describes
the location of the target space where the star product is taken.
It is necessary since $B$ field depends on the target space coordinates.
We note that, historically, the algebra was
realized as the translation group in the presence 
of the magnetic monopole in \cite{Jackiw:1984rd}.

%%% pm %%% 
The phase appearing here has a similar
(but different) form as the phase in the four-point function 
of the open membrane.
However, the assignment of the products are different
between the two viewpoints.
Graphically one may understand it as follows.
The star product between the wave functions of 
the open strings is associated with a triangle where the first
two edges represent the open string wave functions $\phi_{1,2}$
and the third one represents the product $\phi_1\star \phi_2$.
Each side of (\ref{phiphiphi}) contains two products, 
represented by 2 triangles. 
The 4 triangles in this equation form a tetrahedron 
representing the two different
ways to compose three wave functions $\phi_{1,2,3}$,
and it represents the anomaly of associativity. 
The phase appearing in the anomaly is proportional to 
the product of the three momenta.

As we have seen, the phase factor for open membranes 
is also associated with a tetrahedron. 
Four open membranes are placed on
the faces of the tetrahedron, and the tetrahedron
represents the triplet (or the quartic)
product of the wave functions of the open membranes.
It is not directly related to an anomaly in the associativity of the product. 
The phase factor in the exponential is
proportional to {\em the square root} of the product
of the momentum $\vec k_1\cdot (\vec k_2 \times \vec k_3)$.
Nevertheless, we refer to the underlying algebraic structure 
as ``nonassociative'' space in this paper merely because 
the product reduces in the classical (large $C$ field) limit
to a deformation by the Nambu bracket, 
which often hints at a (hidden) nonassociative structure. 
%% pm %%

\subsection{Nambu bracket}
The triplet product which appears in the open membrane
theory is closely related to the Nambu bracket \cite{Nambu:1973qe}.
At the classical level, the Nambu bracket takes the form,
\be \label{NB}
\left\{
\phi_1,\phi_2, \phi_3
\right\}_{\mathrm{NB}}
=
\epsilon_{ijk}\partial_i \phi_1 \partial_j \phi_2 \partial_k \phi_3\,.
\ee
As the Moyal product appears as the quantization of the
Poisson bracket, the product for the open membrane
should be related to the quantization of the Nambu bracket.
It has a long history of the struggle to achieve this goal,
for example, see \cite{Takhtajan:1993vr,
Dito:1996xr,Awata:1999dz,Xiong:2000gp,
Matsuo:2000fh,Kawamura,Curtright:2002fd,
Pioline:2002ba,Berman:2004jv}.
One of the difficulty of the problem is that there seems to be no
natural principle and therefore we have a variety of  
proposals with both merits and demerits.
A natural candidate of the principle is to keep the 
fundamental identities \cite{Hoppe:1996xp},
\bea
\mathrm{(FI-1):}&& \left\{\left\{A_1,A_2,A_3\right\}, A_4, A_5\right\}
+\left\{A_3, \left\{A_1,A_2,A_4\right\}, A_5\right\}
+\left\{A_3, A_4, \left\{A_1,A_2,A_5\right\}\right\} \label{FI1} \\
&&~~~~~-
\left\{A_1,A_2,\left\{A_3, A_4, A_5\right\}\right\}=0\,,\\
\mathrm{(FI-2):}&& \left\{A_{[1},A_2,\left\{A_3, A_{4]}, A_5\right\}\right\}=0\,,\\
\mathrm{(FI-3):} && \left\{\left\{A_1,A_2,A_3\right\}, A_4, A_5\right\}
-\left\{\left\{A_2,A_3,A_4\right\}, A_1, A_5\right\}\nonumber\\
&&~~~~~+\left\{\left\{A_3,A_4,A_1\right\}, A_2, A_5\right\}
-\left\{\left\{A_4,A_1,A_2\right\}, A_3, A_5\right\}=0\,.
\eea
They are the conditions which are necessary that the volume-preserving 
diffeomorphism generated by the Nambu bracket
\be
\delta_{H_1, H_2} x^i = \left\{H_1, H_2,x^i\right\}
\ee
is closed under the commutator. The most difficult issue 
in the quantization of the Nambu bracket comes from
keeping these properties while deforming the bracket.
%%% 0108 %%%
% The sentence above needs to be completed.
%% 0108 %%

At the classical level, there are many ways to implement
the Nambu bracket.
In the open membrane context, it was observed
in \cite{Sakakibara:2000vx, Matsuo:2000fh} that the volume-preserving diffeomorphism
can be generated as the {\em Poisson bracket}
for the closed string on the boundary of the open membrane,
\be
\delta_{H_1, H_2} X^i(\sigma) =\left\{
\omega(H_1, H_2), X^i
\right\}_{PB},\quad
\mbox{ where }\ \omega(H_1, H_2) = \int d\sigma
H_1(X(\sigma))\partial _\sigma 
H_2 (X(\sigma))
\ee
In \cite{Berman:2004jv}, they discussed the spreading
of this closed string into a ``ribbon''
in the presence of the constant three-
form background.  This is analogous to our discussion of
the spreading of the point particle into the open membrane.

The most natural framework of Nambu bracket which
fits with our approach is the one taken in
\cite{Awata:1999dz, Xiong:2000gp, Kawamura}
where the ``cubic matrix'' which generalizes the ordinary
matrix is introduced to define the quantum Nambu bracket.
% The cubic matrix has three indices which can be naturally
% related with the arguments of $\psi(\vec \ell_1, \vec \ell_2 \vec \ell_3)$.
%We write the detail in the next subsection.

\subsection{Connection with cubic matrix}

Here we ``derive'' the product (\ref{tpm})
from the product for cubic matrices
\cite{Awata:1999dz, Kawamura}.
A cubic matrix is an object with 3 indices $A_{lmn}$.
the triplet product for 3 cubic matrices is defined by
\begin{equation} \label{Kawamura}
(ABC)_{lmn} = \sum_k A_{lmk}B_{lkn}C_{kmn}.
\end{equation}
Let us take the limit of infinite dimensional matrices,
and promote the discrete indices to
coordinates of the 3 dimensional continuous space $\mathbf{R}^3$
\be
A_{lmn} \rightarrow A(\vec x_1, \vec x_2, \vec x_3).
\ee
We identify the function $A$ as the wave function
of a triangular closed string, 
and the 3 arguments $\vec x_1, \vec x_2, \vec x_3$
as the position vectors of the 3 points
defining the triangle.

We shall focus on cyclic functions only
\begin{equation}
A(\vec x_1, \vec x_2, \vec x_3) = A(\vec x_2, \vec x_3, \vec x_1),
\end{equation}
where $\vec x_1, \vec x_2$ and $\vec x_3$ are each 3-vectors.

The triplet product (\ref{Kawamura}) now becomes
\begin{equation}
(ABC)(\vec x_1, \vec x_2, \vec x_3) = \int d^3 \vec x
A(\vec x, \vec x_2, \vec x_1) B(\vec x, \vec x_3, \vec x_2) C(\vec x, \vec x_1, \vec x_3).
\end{equation}
This is precisely the product (\ref{tpm}) we used for truncated open membranes
in generic backgrounds.

\end{document}